%% file: ms.tex
\documentclass{emulateapj}

\newcommand {\kms}{km s$^{-1}$}
\newcommand {\ang}{\AA\ }
\newcommand {\angns}{\AA}

\newcommand {\lya}{Ly$\alpha$ }

\newcommand {\lyb}{Ly$\beta$ }
\newcommand {\lyc}{Ly$\gamma$ }
\newcommand {\lyd}{Ly$\delta$ }
\newcommand {\lye}{Ly$\epsilon$ }
\newcommand {\lyf}{Ly$\zeta$ }
\newcommand {\lyg}{Ly$\eta$ }
\newcommand {\lyh}{Ly$\theta$ }

\shorttitle{Lyman-$\alpha$ Forest of PC 1643+4631A,B}
\shortauthors{C.~M. Casey et al.}

\begin{document}
  
  \title{PC 1643+4631A,B: THE LYMAN-$\alpha$ FOREST AT THE EDGE OF COHERENCE}

  \author{C. M. Casey\altaffilmark{1,2}, C. D. Impey\altaffilmark{1}, C. E. Petry\altaffilmark{1}, 
    A. R. Marble\altaffilmark{1}, R. Dav{\'e}\altaffilmark{1}}
  \altaffiltext{1}{Steward Observatory, University of Arizona, Tucson, AZ 85721 U.S.A.}
  \altaffiltext{2}{Institute of Astronomy, University of Cambridge, Madingley Rd, Cambridge, CB3 0HA, U.K.}

  \begin{abstract}
  This is the first measurement and detection of coherence in the intergalactic medium (IGM) at 
  substantially high redshift (z$\sim$3.8) and on large physical scales ($\sim$2.5 $h^{-1}_{70}$ Mpc).  
  We perform the measurement by presenting new observations of the high redshift quasar pair 
  PC 1643+4631A, B and their \lya absorber coincidences.  With data collected from Keck I Low 
  Resolution Imaging Spectrometer (LRIS) in a 10,200 sec integration, we have full coverage  
  of the \lya forest over the redshift range $2.6 < z < 3.8$ at a resolution of 3.6\ang ($\sim$ 
  220 \kms).  This experiment extends multiple sightline quasar absorber studies to higher 
  redshift, higher opacity, larger transverse separation, and into a regime where coherence 
  across the IGM becomes weak and difficult to detect.  Noteworthy features from these spectra 
  are the strong Damped \lya Absorbers (DLAs) just blueward of both \lya emission peaks, each 
  within 1000 \kms\ of the emission redshift but separated by 2500 \kms\ from each other. The 
  coherence is measured by fitting discrete \lya absorbers and by using pixel flux statistics. 
  The former technique results in 222 \lya absorbers in the A sightline and 211 in B.  Relative 
  to a Monte Carlo pairing test (using symmetric, nearest neighbor matching) the data exhibit a 
  4$\sigma$ excess of pairs at low velocity splitting ($\Delta v < 150$ \kms), thus detecting 
  coherence on transverse scales of $\sim$2.5 $h^{-1}_{70}$ Mpc.  We use spectra extracted from 
  an SPH simulation to analyze symmetric pair matching, transmission distributions as a function 
  of redshift and compute zero-lag cross-correlations to compare with the quasar pair data.  The 
  simulations agree with the data with the same strength ($\sim$4$\sigma$) at similarly low 
  velocity splitting above random chance pairings.  In cross-correlation tests, the simulations 
  agree when the mean flux (as a function of redshift) is assumed to follow the prescription given 
  by \citet{kirkman05a}.  While the detection of flux correlation (measured through coincident 
  absorbers and cross-correlation amplitude) is only marginally significant, the agreement between
  data and simulations is encouraging for future work in which even better quality data will provide 
  the best insight into the overarching structure of the IGM and its understanding as shown by SPH 
  simulations.
  \end{abstract}
  

\section{INTRODUCTION}

Quasars have become vital tools for understanding the nature of the intergalactic medium (IGM) 
over most of the Hubble time.  Blueward of \lya emission, the absorption in the \lya forest 
traces the one-dimensional distribution of the neutral component of the intervening material$-$the 
intergalactic medium$-$along the line of sight.  Since the first analyses of the forest by 
\citet{sargent80a}, astronomers have used \lya absorption spectra to characterize 
the physical state of the IGM at various epochs. The IGM is highly ionized and absorption of 
neutral hydrogen (\ion{H}{1}) scales as a power law according to the gas density \citep{rauch98a}.  
At the lowest redshifts, this so-called Gunn-Peterson approximation breaks down since a non-negligible
amount of the absorbing hydrogen has collapsed into dense structures or warm narrow-line \lya
absorbing gas. 

Over the past twenty years, several studies have shown that \lya absorbers 
are better tracers of dark matter potential wells and baryon density than galaxies,
supporting expectations from theory and consistent with gas dynamic simulations 
\citep{cen94a,zhang95a,hernquist96a,dave99a}.  The HST Quasar Absorption Line Key
Project \citep{bahcall96a,jannuzi98a,weymann98a} used many single sightlines
with $z \le 1.6$ to probe the complex geometry of the evolving ``Cosmic  Web'' 
through its voids and dense regions.  The role of hydrodynamic simulations becomes 
important when considering bulk properties of the IGM as inferred by \lya flux
decrements \citep{marble07a}; the data from the \lya forest show the characteristics 
of the universe in long but widely-separated redshift paths while simulations perform 
the forward experiment, modeling the evolution of the universe in finite volumes that 
are limited by number of particles and resolution.

Multiple sightline experiments, originally presented using gravitationally lensed 
quasars \citep{shaver82a, shaver83a, weymann83a, foltz84a, smette92a, smette95a}, 
are used to infer the IGM's structure in two dimensions through cross-correlation 
of the spectra giving characteristic coherence lengths.  With pairs (or groups)
of different transverse separations and different redshifts, the evolution in the
typical structures of the IGM can be measured.  The measurement of coherence is based 
on the statistical excess of matched, coincident absorber features across the lines 
of sight.

The motivation for extending coherence measurements to higher redshift
and larger transverse separation, where the signal is anticipated to be
weak, is that it represents a new regime for testing the gravitational
instability paradigm that underlies the description of large scale 
structure on scales larger than galaxies.  One study yielded a marginal
detection of line correlations in a grid of 10 quasars on transverse scales 
of $\sim 10 h^{-1}_{70}$ at $z \sim 2.5$ \citep{williger00a}, but the authors
did not compare the signal to theorectical expectations.  The same group
subsequently used moments of the transmission probability density to measure
the coherence \citep{liske00a} and saw diagreement with the hydrodynamic
simulations of \citet{cen97a}, although they did not make direct comparisons
as in this paper.  Other motivations for this type of study include the
possibility of discovering non-Gaussian structures like voids, as observed
on these scales at $z \sim 2$ by \cite{rollinde03a}, or testing for 
non-gravitational effects, as anticipated by \cite{fang04a}.

Line of sight correlations are washed out on distance scales less than the
the redshift resolution, and the transmission distribution on velocity
scales of less than several hundred kms$^{-1}$ is modulated by peculiar
velocities.  Transverse correlations can be effectively measured given a
mean $SNR > 10$, which allows peaks in the opacity distribution to be 
located in velocity reliably and with a precision better than instrumental
resolution.  Suitable interpretation of course depends on treating data and
hydrodynamic simulations identically for the coherence measurement.

These observations of the high redshift quasar pair PC 1643+4631A, B focus on the redshift 
range from $z = 2.6$ (Lyman limit cutoff at 4377\angns) to \lya emission 
at $z_{em} = 3.8$.  With better resolution and higher signal-to-noise than previous 
William Hershel Telescope (WHT) spectra analyzed by \citet{saunders97a}, these data 
allow a detailed study of the \lya absorbers.  \citet{saunders97a} 
had 12\ang instrumental resolution and a 1500 sec integration yielding poor
blue data sufficient only to determine if the pair was a lens, since there had been
earlier claims of a strong CMB decrement towards the pair \citep{jones97a}.
They did not present analysis of \lya absorber coincidences.  At redshift 3.790, quasar 
A has magnitude $B_{A}$ = 20.0 $\pm$ 0.3 and quasar B, at redshift 3.831, has magnitude 
$B_{B}$ = 20.7 $\pm$ 0.3; the two quasars have an angular separation of $198\arcsec$.  
The spectra and redshifts are sufficently different that there is no doubt that the 
quasars form a physical pair rather than a lens system.

The best basis for estimating the requirement for a definitive detection of coherence on 
3 $h^{-1}$ Mpc scales comes from the work of \citet{rollinde03a}.  They observed five pairs 
and a quad spanning angular scales that encompass the anglular separation of the pair in this 
paper, and their VLT data has almost identical resolution. Their N-body simulations indicate 
that a mean $SNR \sim 50$ yields a 1$\sigma$ detection of flux correlation for a separation 
of 200 arcseconds, so assuming Gaussian noise, a 3$\sigma$ detection would take 10 pairs with 
this quality of data or a single pair with $SNR \sim 150$.  In this experiment we explore two
methods of measuring coherence; the first fits the predictions of \citet{rollinde03a} since 
measuring coherence is difficult with poor SNR$\sim$30, and the second is matching \lya absorbers 
across the sightlines using nearest neighbor matching.

In \S \ref{obsdr_s} we discuss the observations and data reduction, and \S \ref{g6sims_s} 
introduces the mechanics of obtaining spectra from simulations.  The use of automated 
software, named $ANIMALS$, to fit lines and continua to spectra and simulations is described 
in \S \ref{animals_s}.  Physical characteristics of the spectra are given in \S \ref{charspec_s}, 
while the absorber characteristics are given in \S \ref{lines_s}.  The coincidences between 
sightlines are described in \S \ref{linecoincide_s}.  We compare simulation extractions to 
the data in \S \ref{sph_s}, both in terms of individual sightlines and their transmission 
properties, and jointly to understand coherence in the transverse dimension.

\section{OBSERVATIONS AND DATA REDUCTION}\label{obsdr_s}

\subsection{Processing the LRIS Spectra}\label{lris_ss}

All data for this project were obtained with the Keck I Low 
Resolution Imaging Spectrometer (LRIS) on 21 July 2001 and were reduced 
using the standard $IRAF$ longslit-spectra pipeline.  To cover the wavelength 
range of scientific interest, only the red side of the spectrograph 
and the 900/5500 grating was used.   The red CCD has two amplifiers; at
the time of the observations the left side had a gain of 1.97 $e^{-}$/ADU and 
read noise 6.3 $e^{-}$, and the right had gain 2.10 $e^{-}$/ADU and read noise 6.6 $e^{-}$.
The LRIS red CCD also has unexposed rows due to the sub-array readout, but
these presented no practical problem in the data analysis.  Since both 
quasars were placed on the long slit simultaneously, all images
were examined to make sure no alignment shift occurred during the observations.
The target was continuously tracked and the slit was aligned close enough to
the parallactic angle to avoid significant loss of blue light. 
Lyman limit absorption means there is little useful data below 4400\angns.  These July
2001 data had an effective integration time of 10,200 seconds.  A previous
dataset from 2 June 2001 was of too poor quality to be useful.

While most data reduction steps were performed with the standard $IRAF$ packages,
we used two tasks ($lccdproc$ and $lrisbias$) that are specific to LRIS.  The 
column bias effects from the overscan region were removed from all science 
frames using $lccdproc$ with gain of 1.97 and read noise 6.3 $e^{-}$ entered as
parameters for the left channel.  The images were then trimmed to the appropriate
size and $zerocombine$ was used to combine the bias frames.  The biases were
averaged together to reduce cosmic rays with a sigma-clipping algorithm.  
With $lccdproc$, this averaged bias frame was then subtracted from the remaining
images.

To remove pixel-to-pixel variations and optical vignetting, we created a flat
field image using $flatcombine$.  The only complication was a gain discontinuity
between right and left amplifiers. The discontinuity is not noticeable in 
the low exposure frames (the science images), but it was removed by dividing
each side of the combined flat by its respective gain.  Once this correction
had been made, we used $response$ to fit a continuous spectral response function
to the flat frame.  After the resultant flat field was normalized, the remaining
images (flux calibration and science) were divided by the flat field, removing
pixel-to-pixel variations down to a level of 3$\%$.

\subsection{Spectral Extraction}\label{specex_ss}

For cosmic ray rejection, we combined five individual 1800 second 
exposures and one 1200 second exposure in 2D before extraction.  To line up 
the quasar spectra, each image was shifted in x and y coordinates using $imshift$.  
The images were then averaged together using $imcombine$ and its 'cosmic ray
rejection' algorithm.  Prior to extracting the spectra using $apall$ (part of 
the package $APEXTRACT$), suitable regions for sky subtraction were chosen by 
visual inspection. There was a star close to quasar B that had to be avoided.  
For quasars A and B, the trace had a FWHM of 6-7 pixels, indicating seeing 
of 0.6\arcsec\ during the observations.  Even after sky subtraction, the strong sky line at 
5577\ang shows up as a residual artifact in the final spectra. $Apall$ was used to create 
a weighted variance spectrum array, an unweighted spectrum array, a background 
spectrum array, and an error array for each quasar.  A second order trace function 
was used to linearize the spectra.  

\subsection{Wavelength and Flux Calibration}\label{svfxcalib_ss}

For wavelength calibration, the lamp spectra were extracted using the same 
trace functions given by the quasar pair A and B.  HgNeArCdZn and CdZn lamps 
were taken before and after the quasar exposures.  The $IRAF$ tasks $identify$
and $reidentify$ were used to iteratively fit lines, with solutions for all 
permutations of the two lamps and the two target apertures inspected.  The 
final wavelength solution applied to A and B was based on the HgNeArCdZn 
lamp since it had a more uniform distribution of lines; after excluding one line 
from the fit, the final RMS wavelength error was 0.054\angns, which is $\sim$2$\%$ of 
a resolution element.  The wavelength solution was applied using the $IRAF$ tasks 
$refspec$ then $dispcor$ sequentially.  The sampling of the final reduced spectra 
was 0.85\ang per pixel.

\begin{figure*}
  \centering
  \includegraphics[width=1.90\columnwidth]{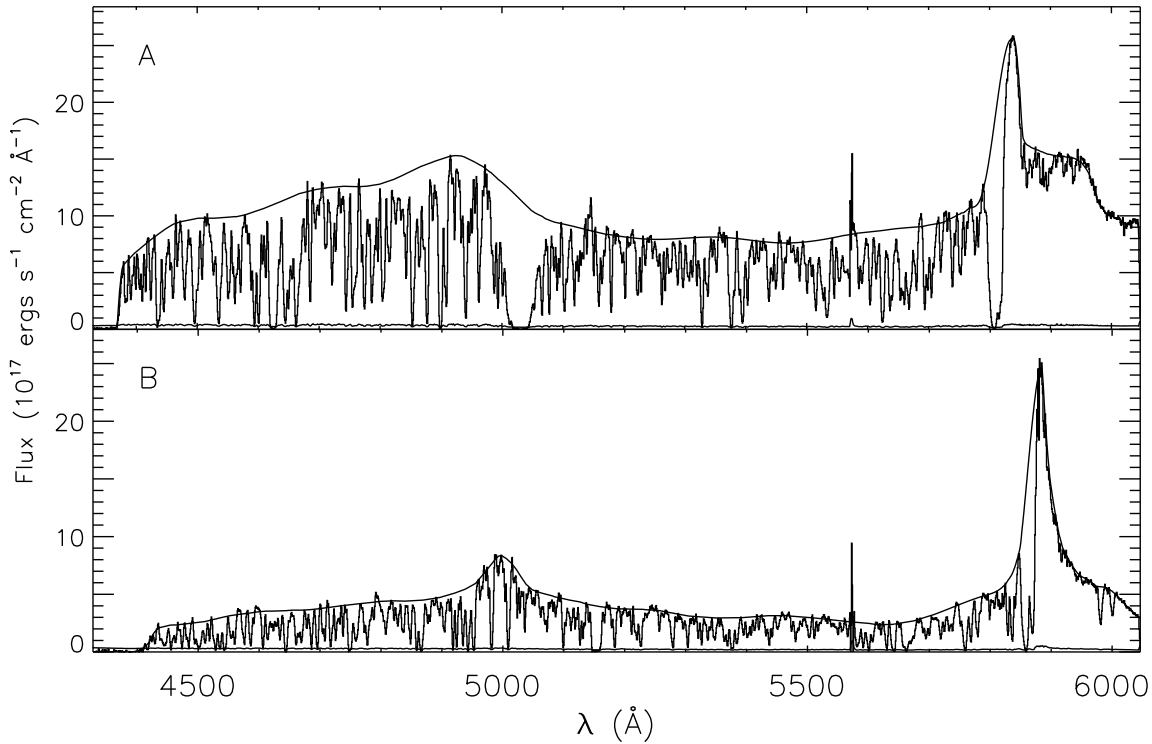}
  \caption{
    The spectra of the quasars PC 1643+4631A (top) and B (bottom).  The quasars have
    magnitudes $B_{A} = 20.0$ and $B_{B} = 20.7$, and the mean signal to noise ratios are 32 and 16 respectively.
    The red end of the LRIS spectrograph covers wavelengths 4200\ang to 6100\angns, yielding the full
    coverage of the \lya forest, but little data redward of \lya emission.  
  }
  \label{allspec_f}
\end{figure*}

Flux calibration is not necessary to achieve the scientific goals of the 
experiment, but it is useful for seeing the true continuum shape.   To 
calibrate the flux, we divided each amplifier by its gain (to eliminate the 
right/left channel discrepancy) and remultiplied by 1.97 (the left channel 
gain).  This preserves the normalization of the counts, but eliminates the 
discontinuity of the spectrum.  We ran $apall$ to extract a spectrum for the 
standard star, BD 284211, then used $standard$ and $sensfunc$ to fit the sensitivity 
function of the LRIS instrument.  The flux calibration was applied using 
$calibrate$ and the final spectra are shown (with 1.5 pixel boxcar smoothing) in Figure
\ref{allspec_f}. The flux and error arrays are available in the electronic version of 
the article.

\subsection{Instrumental Resolution and Signal to Noise}\label{irsnr_ss}

To calculate a representative value for instrumental resolution, we used the 
average FWHM from our observation run.   Blended lines, saturated lines, 
peripheral lines widened by defocusing, and low intensity lines were excluded 
from the resolution calculation, and only the FWHM of lamp lines in the \lya
forest wavelength range were used.  Averaging the FWHM of 11 features with
characteristic Gaussian line profiles gives an instrumental resolution of 
3.6 $\pm$ 0.3\angns.  At 5000\angns, this corresponds to $\sim$220 \kms.

To calculate signal to noise (SNR) for each quasar, we used the output flux and 
error arrays from $IRAF$.  First, we divided the signal by the error for each 
wavelength bin and averaged the SNR value over the \lya forest range. 
This method gives the SNR of quasar A as 32.2 $\pm$ 5.7 and 16.8 $\pm$ 4.3 for 
quasar B.  A second method averages the signal over the \lya region (fairly 
constant), averages the flux error similarly, and divides the two values.  
This gives the SNR for A as 31.3 $\pm$ 7.4 and SNR for B as 16.6 $\pm$ 6.4.  
Since the methods agree well, we round the SNR values to 32 for quasar A and 
16 for quasar B in the remainder of this paper.

\section{SPECTRA DRAWN FROM SIMULATIONS}\label{g6sims_s}

We will compare our cross-correlation measurements with simulations, to see if
current structure formation models can reproduce observations.  For this purpose, 
we considered pairs of synthetic Ly$\alpha$ absorption spectra extracted from the 
smoothed particle hydrodynamic (SPH) cosmological simulation G6 \citep[described 
below;][]{springel03a,finlator06a}. We extracted 1,000 paired lines of sight (separated
by 198 arcseconds, converted to physical transverse distance as a function of 
simulation redshift) randomly distributed across each of the three orthogonal 
faces of the simulation box for four different epochs ($z=2.5,\ 3.0,\ 3.5,$ 
and $4.0$) giving a total of 24,000 individual spectra. The physcial transverse 
separations of sightlines at these four epochs are 2.16, 2.54, 2.61, and 2.66 
$h^{-1}_{70}$ Mpc respectively.  These extractions were originally made for a 
study of the Alcock-Paczynski effect in the Ly$\alpha$ forest, and are described 
in more detail in \citet{marble07a}.

The G6 simulation is an extension of the G-series \citep*{springel03a}, which was 
run with a modified version of the N-body+SPH galaxy formation code {\sc Gadget} 
\citep{springel01a}.  The following cosmological parameters were employed:
$\Omega_m=0.3$, $\Omega_\Lambda=0.7$, $\Omega_b=0.04$, $\sigma_8=0.9$, and $h=0.7$, 
consistent with first-year WMAP data as well as Ly$\alpha$ forest power spectrum 
results \citep{spergel03a}.  The relatively large volume of G6 (100 $h^{-1}$ Mpc 
comoving on a side) is necessary to accurately model matter correlations on the 
scale of few Mpc.  With $486^3$ dark matter and $486^3$ gas particles, it is one 
of the largest cosmological hydroydnamic simulations ever done, and its large 
dynamic range is critical for resolving the small-scale structure associated with
the Ly$\alpha$ forest.  The corresponding gas mass resolution  of $9.79\times10^7$ 
$h^{-1}M_{\sun}$ is not ideal for detailed study of the Ly$\alpha$ forest as it does not 
quite resolve the Jeans mass in the IGM \citep{schaye01a}, but given that the observations 
considered here have a spectral resolution of 220~km/s (i.e. a distance of about 
0.7~Mpc/h in Hubble flow), the mean interparticle spacing of 0.2~Mpc/h is sufficient 
to resolve fluctuations at the level probed by the data.

Radiative heating and cooling were calculated assuming photoionization equilibrium and 
optically thin gas. Additional prescriptions for star formation and supernova feedback 
were incorporated, though the impact of these subgrid prescriptions on the Ly$\alpha$ 
forest is minimal.  Additionally, G6 included galactic outflows via a Monte Carlo 
ejection of gas from star-forming regions, where the outflow speed is taken to be 
484~km/s and the mass loading factor (i.e. the mass outflow rate relative to the star 
formation rate) is taken to be 2.  Recent work has shown that this particular outflow model
is not as successful as one where the outflow velocity scales with the characteristic 
velocity of the galaxy \citep[i.e. the ``momentum-driven wind'' scalings of][]{oppenheimer06a}.  
For instance, momentum-driven wind scalings better reproduce IGM enrichment~\citep{oppenheimer06a},
the galaxy mass-metallicity relation~\citep{finlator08a}, and observations of high-redshift 
galaxies~\citep{dave07a}, among other things.  However, \citet{marble07a} demonstrated that 
the outflow prescription has only a very minor impact on correlations in the Ly$\alpha$ 
forest; in particular the difference in Ly$\alpha$ forest correlation between using the 
G6 wind model and a momentum-driven wind model is $\scriptstyle\lesssim$ 5\%.  This is 
primarily because outflows enrich a relatively small volume of the Universe~\citep{oppenheimer06a}, 
so do not impact the bulk of the volume traced by Ly$\alpha$ absorbers.

Finally, a spatially uniform photoionization background was included, with
the spectral shape and redshift evolution given by \citet*{haardt96a}.
For moderate overdensities characteristic of the Ly$\alpha$ forest
($\rho/\langle\rho\rangle < 10$), the amplitude of this background can be
effectively and subsequently changed by rescaling the resulting opacity
distribution \citep{croft98a} to match the desired mean transmitted flux,
$\langle f \rangle$, of the Ly$\alpha$ forest.  However, $\langle f \rangle$ 
remains an observationally uncertain quantity.  Measurements of 
$\langle f \rangle$ made directly from high-resolution echelle spectroscopy 
(\emph{e.g.}, \citet*{kim01a} and \citet{kirkman05a})  have yielded consistently 
higher values than those based on extrapolation from redward of the Ly$\alpha$
forest in much larger samples of lower-resolution spectra (\emph{e.g.},
\citet*{press93a} and \citet{bernardi03a}), with relative differences of
$5\%-25\%$ at $2.5<z<4.0$.  For this reason, two sets of simulated spectra
were produced, corresponding to the mean flux prescriptions of both
\citet*{press93a} and \citet{kirkman05a}, hereafter referred to as P93 and
K05.

\section{$ANIMALS$ PROCESSING}\label{animals_s}

To detect, fit and analyze quasar absorption lines, we used a custom made software package: ANalysis 
of the Intergalactic Medium via Absorption Lines Software ($ANIMALS$), originally based on a kernel 
developed by Tom Aldcroft with substantial modifications by Cathy Petry at Steward Observatory.  
The software is faithful to the methodology of the HST Quasar Absorption Line Key Project 
\citep{bahcall96a,jannuzi98a,weymann98a}, and its characteristics are described at length in the 
original method paper \citep{petry98a}.  While it was originally designed to handle typical observed 
quasar spectra, it has since been expanded to accomodate spectral extractions from SPH simulations 
(which have much shorter spectral coverage but higher resolution than is typical of real observations)
and automate the process of continuum fitting \citep{petry02a}.  This expansion also included improvements 
to handle data with a wide range of SNR and line density$-$particularly useful when analyzing high 
redshift quasar \lya forest lines.  For a more complete discussion of the software, see \citet{petry06a}.

We use $ANIMALS$ in three contexts in this paper$-$to fit continua to the data, to fit absorption lines and 
measure their significances and equivalent widths (EW), and to degrade spectra of SPH extractions to the 
instrumental resolution and signal-to-noise of our observations.  These processes are described below, 
and their purposes become apparent later in the paper.

\subsection{Continuum Fitting}\label{contfit_ss}

The continua for both PC 1643+4631A and B were fit by automatically computing the average 
flux value in 50\ang  bins using $ANIMALS$'s simple cubic spline fitting routine. 
If points on the spectrum deviated by more than $2\sigma$ from the averaged flux values 
(the ``continuum'') in the negative direction, they were flagged as potentially absorbed pixels
and rejected from the averages that tether subsequent iterations of the 
continuum fit.  The procedure generally converged after four iterations, after 
which the spline fit rested near the top of the spectral features.

This initial fit was not completely adequate.  Continuum fitting is made more 
challenging by the high redshift of the quasars and modest spectral resolution of the data, which means 
there is \ion{H}{1} opacity at essentially every pixel. Combined with the high signal-to-noise 
ratio, this implies that most of the small scale structure is real transmission variation.  
Based on the structure of the quasar continuum and careful scrutiny of each 
feature, the continuum was 
adjusted--with a stiffer fit in most cases (on 80\ang scales instead of 50\ang scales), or (near the \lya or \lyb 
emission lines) a higher order fit.  In a few regions final manual adjustment 
had to be made to the spline fit, where individual points were connected with 
straight lines on a scale of 20 \angns. The result is a continuum that sits at 
or just below spectral peaks (less-absorbed regions) and is low-order enough not 
to follow individual absorber features or fluctuations due to noise.

We also used $ANIMALS$ to fit continua to our SPH extractions to analyze the reliability
of our continuum fitting techniques for different line densities and underlying opacity distributions
as a function of redshift.  The results are given in \S \ref{underestimation_ss}, but 
in \S \ref{specdeg_ss} we discuss minor differences between fitting the simulations and the observed data.

\subsection{Line Fitting}\label{linefit_ss}

$ANIMALS$ fits Gaussian profiles to individual absorbers in regions where the flux is below the continuum fit.
The line fitting algorithm allows the central wavelength, amplitude and FWHM of each line to be variables.  
While no limit is set to the fitted equivalent width ($W_{obs}$), this analysis is based on the assumption 
that individual features are unresolved, since the spectral resolution ($\sim$220 \kms) is well beyond the 
upper bound of the distribution of absorber Doppler parameters, as measured by echelle observations \citep*{kim01a}.  
At the redshifts of this experiment, 80$\%$ of the absorbers will have Doppler parameters in the range 20-50 
\kms, contributing a negligible broadening in quadrature to the instrumental profile. The damped features 
described in \S \ref{charspec_s} are excised from the spectra during line fitting since $ANIMALS$ is not designed 
to fit the profiles of high column density absorbers.

\input{tab1}
\input{tab2}

The process of line selection in $ANIMALS$ focuses on maximizing real features and reducing 
contamination from false signals$-$caused by noise or variation in the true continuum, which 
is unknown a priori \citep{petry98a}.  Applying the software to PC 1643+4631A and B yields 240 
and 234 unresolved fitted lines respectively. A list of all the detected lines is found in Tables 
\ref{tab1} and \ref{tab2}, where each line is characterized by its central wavelength, observed 
$EW$ ($W_{obs}$), $\chi^{2}_{\nu}$ (based on the quality of the fit), $S_{fit}$ and $S_{det}$ 
(significances to be discussed in \S \ref{sigs_ss}), and special notes.  Identification of the 
\lya sample is explained in \S \ref{lyasample_ss}, resulting in 222 A lines and 211 B lines that 
are used for the analysis.  We also perform line fitting of SPH simulation spectra, whose details
are discussed in \S \ref{simlines_ss}.

\subsection{Spectral Degradation}\label{specdeg_ss}

For a fair comparison with observations, the spectra extracted from the SPH simulation 
must be degraded to the instrumental resolution and observed SNR of the observations.
To degrade, the spectra were convolved with a Gaussian line spread function, resampled 
to the correct dispersion (0.85\ang per pixel) using spline interpolation, and Gaussian 
noise was added to match the SNR as a function of wavelength given by the data. 

Since the simulation extractions each span only 300\ang (corresponding 
to a 100 $h^{-1}_{70}$ Mpc box), automatic continuum fitting can be problematic near 
the boundary.   Although periodic boundary conditions in the simulation 
prevent any discontinuity in transmission at the box edge, an ``unlucky''
anisotropy in opacity, or region of heavy absorption, can skew or tilt the continuum 
fit.  To avoid large transmission deviations in the continuum 
fit, the simulated spectra were replicated three times using a different 
noise seed and joined end-to-end.  Similar to the continuum fits for the data, $ANIMALS$ 
was used to automatically fit continua to the degraded simulation extractions using a 
cubic spline fitting routine. The fits used 170-255\ang 
smoothing, and the middle third of each fit was adopted as the continuum.

\section{CHARACTERISTSCS OF THE SPECTRA}\label{charspec_s}

\subsection{Spectral Features}\label{specfeat_ss}

\begin{figure}
\includegraphics[height=0.99\columnwidth,angle=270]{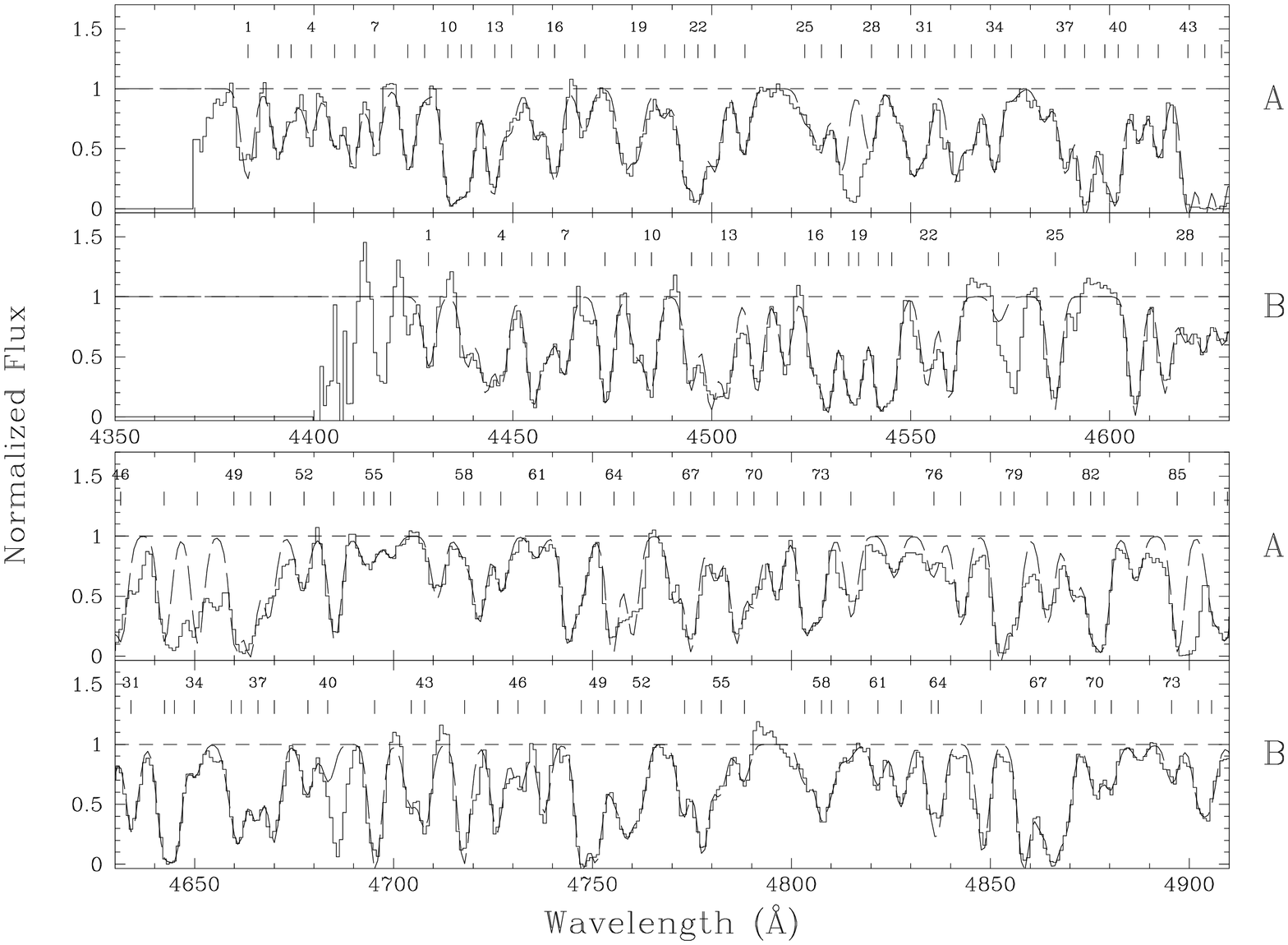}
\includegraphics[height=0.99\columnwidth,angle=270]{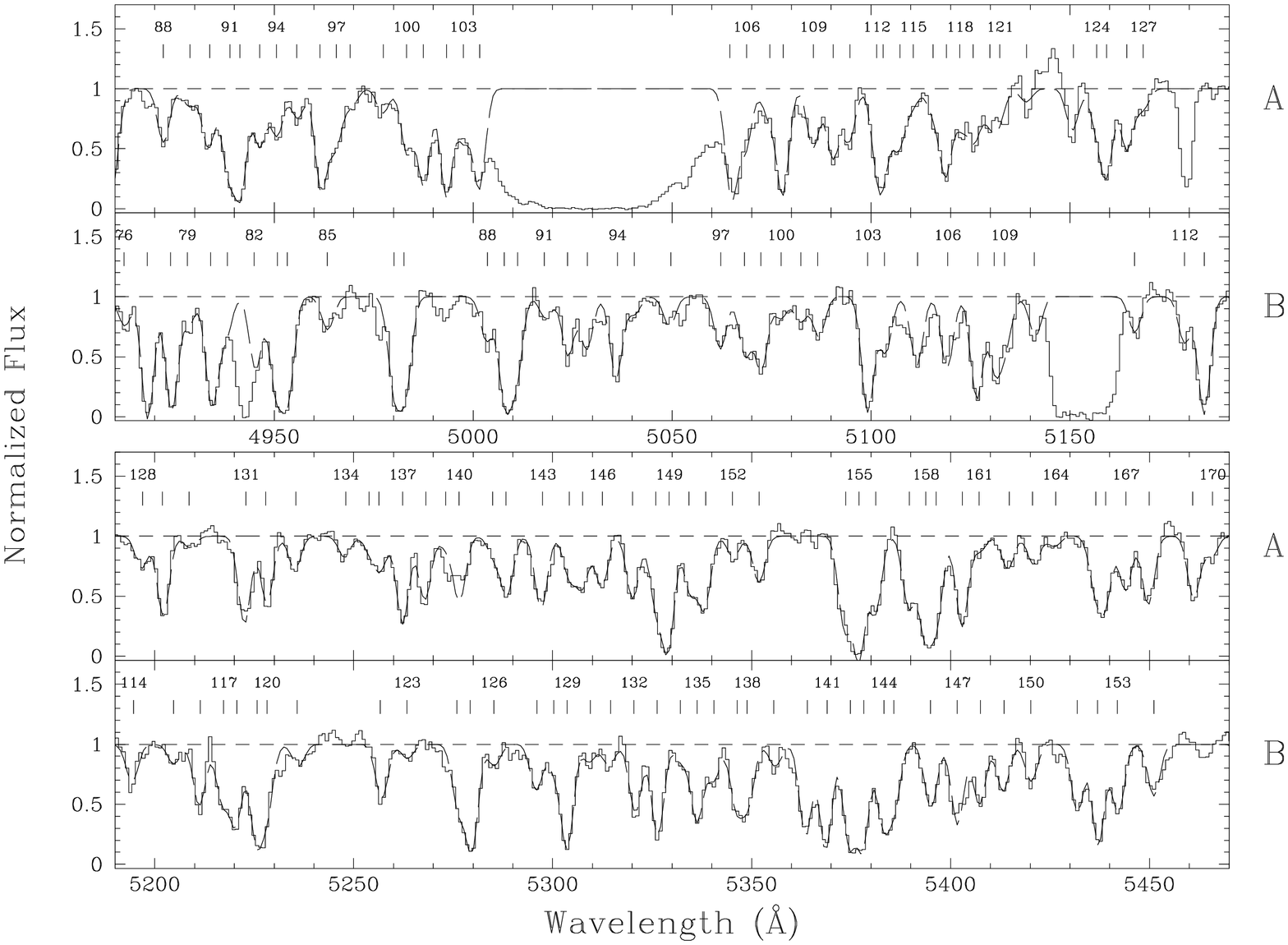}
\includegraphics[height=0.99\columnwidth,angle=270]{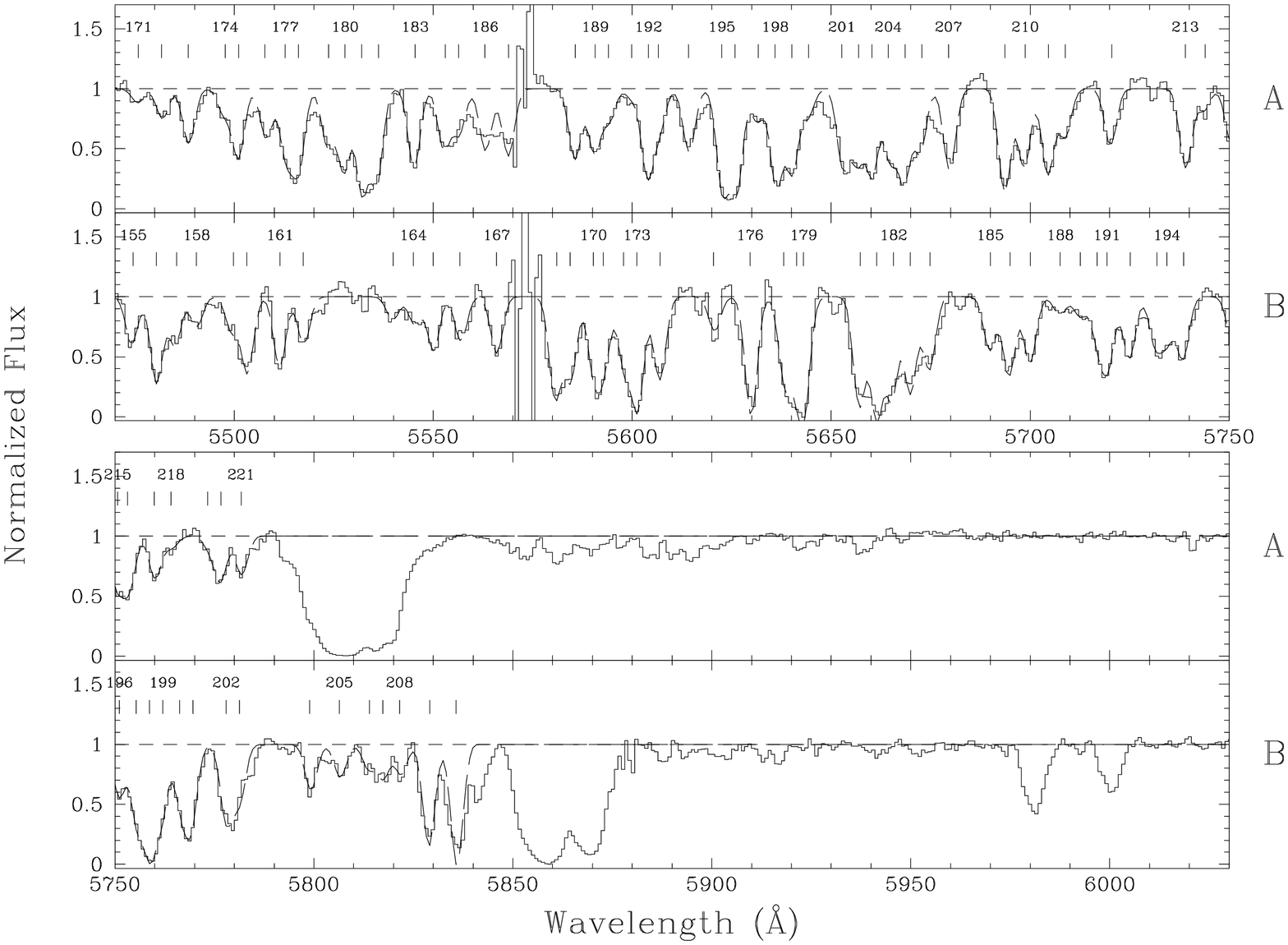}
\caption{
  The normalized flux (relative to the continuum fit) of the spectra of quasars PC 1643+4631A (above)
  and B (below).  Fitted absorption lines from the \lya sample are numbered (every third line) above 
  the spectra and are traced by the dashed line overlayed on the spectra.  For wavelengths below 4370\ang
  in spectrum A and below 4400 in spectrum B, the flux is nearly zero due to Lyman limit absorption,
  so the normalized flux in these regions has been set to zero.
}
\label{expandspec_f}
\end{figure}

The quasar spectra in Figure \ref{allspec_f} show \lya emission peaks at wavelength 5838\ang 
for A and 5880\ang for B.  An expanded normalized flux plot of the spectra is shown in Figure 
\ref{expandspec_f}. These \lya wavelengths are redward of the values anticipated from the published 
redshifts \citep{schneider91a} at 5823\ang and 5872\angns, but their peaks are possibly shifted due
to heavy absorption by strong intervening damped Lyman absorbers (DLAs) just blueward of \lya emission.   

Both spectra exhibit strong, independent damped Lyman absorbers just blueward of \lya emission at 5808\ang 
in spectrum A and at $\sim$5859\ang in B.  The DLAs have separations of 770 \kms\ and 665 \kms\ from their 
respective quasars (A and B) and are thus associated systems.  The absorption in spectrum B is bimodal and 
consists of two different absorbers: one at $z = 3.827$ which is intrinsic to the quasar with strong 
\ion{N}{5} absorption, and another at $z = 3.818$ (at 5859\angns).  We claim that two independent absorbers,
at $z = 3.783$ (5815\angns) in A and at $z = 3.818$ (5859\angns) in B, are DLAs.  The occurrence of a single 
close associated DLA with a quasar is quite rare ($\sim$2$\%$), so the presence of two of these features in 
a pair with fairly large separation is significant \citep{hennawi06a}.  The DLAs are too widely separated 
from each other (by 2500 \kms\ or $\Delta z = 0.012$) to be considered part of the same cosmic structure, 
showing a velocity separation roughly the same as that between the quasars.  These absorbers, and the quasars 
themselves, may probe a much larger overdense region at $z \sim 3.8$.

In spectrum A, we see another strong damped feature from 5005-5060\ang.  This region is blocked out of 
\lya absorber analysis.  In spectrum B, there is a somewhat weaker system ranging from 5145-5160\angns.  
Both spectra show Lyman limit systems, at  4370\ang in A and at 4409\ang in B.  Redward of \lya there is 
very little wavelength coverage, but a strong \ion{N}{5} doublet appears in spectrum B at redshift 3.82.  
No lines redward of \lya could be identified.

\subsection{Redshift Estimation}\label{zguess_ss}

The most recent estimations of the pair's redshift, from \citet{schneider91a}, is $z=$3.790 $\pm$ 0.004 for A and $z=$3.831 $\pm$
0.005 for B.  Since neither spectrum presented here has significant redward coverage, any 
re-estimation of the redshifts must be done with \lya emission, \lyb emission, and the corresponding 
Lyman limit.  The spectrum for quasar A is too irregular to reliably pick out \lyb emission, the peak of 
\lya emission is slightly shifted as described previously, and the Lyman limit (while agreeing with the
published redshift) cannot give a very precise redshift estimate.  In spectrum B, 
\lyb emission is well-traced by the continuum fit, but the fitted peak is located 40\ang
redward of the expected \lyb emission peak (using the published redshift z = 3.831 $\pm$ 
0.005).  While this offset might 
be attributed to a faulty wavelength solution, we found that both the \lya and the Lyman limit agree
with the published redshift, and the 5575\ang sky line affirms that there is no 
zero point shift in the wavelength solution. The deficit blueward of the \lyb peak is attributed 
to absorption associated with the DLA system seen just blueward of \lya emission.  This, 
coupled with OVI emission at 1036\angns, creates the impression of the shifted \lyb peak. Without any
sharp and non-absorbed emission peaks, we confirm but cannot improve on the published redshifts.

\section{CHARACTERISTICS OF THE ABSORBERS}\label{lines_s}

The line-fitting methodology is taxed at high redshift because increased opacity significantly 
affects the reliability of line fitting. High opacity can cause a substantial underestimation of
continuum flux and thus underestimation of line density and equivalent width.  At high redshift, 
line densities are high and blending is a severe problem.  We discuss 
these effects quantitatively with SPH simulation extractions in \S \ref{underestimation_ss}.  
Below we describe the characteristics of absorbers, define the \lya sample, discuss 
contamination from metals and higher order Lyman lines, and calculate line densities.

\subsection{Line Significances}\label{sigs_ss}

\begin{figure}
  \includegraphics[width=0.99\columnwidth]{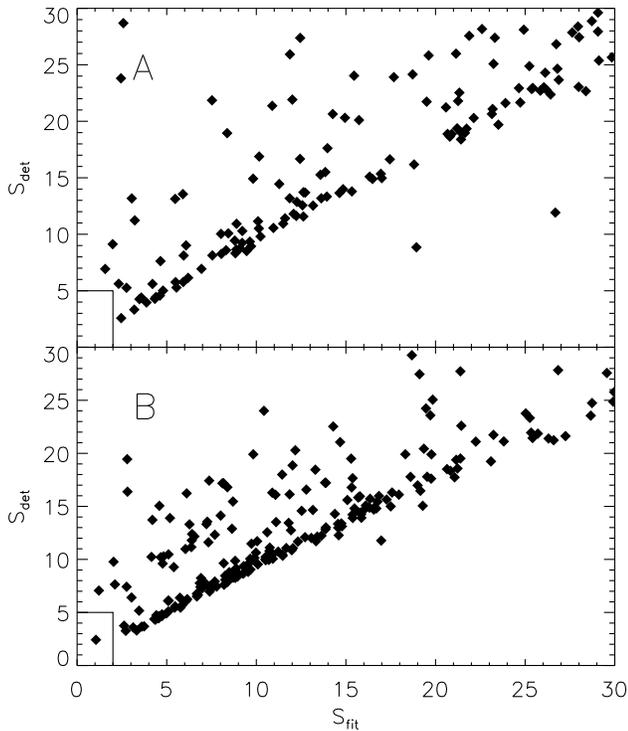}
  \caption{
    Fitted significance vs. detection significance for A and B absorbers  (76 highly 
    significant lines from A and 9 from B lie outside of the plot boundaries).
    Based on previously determined significance parameters discussed in 
    \S \ref{sigs_ss}, we exclude lines with $S_{fit} < 2$ and $S_{det} < 5$ from the 
    \lya sample, only containing one line in either spectra: a blended line at 5114\ang in B.
  }
  \label{sigs_f}
\end{figure}

ANIMALS provides two independent measures of absorption line
identification reliability.  The detection significance, $S_{det} =
W_{obs} / \sigma_{det}$, relates the strength of each line to the
detection limit of the data at the corresponding wavelength.  Here,
$\sigma_{det}$ is given by the convolution of the instrumental line
spread function with the 1$\sigma$ flux error array, and, as before,
$W_{obs}$ is the fitted equivalent width.  How well a given line is
fit is quantified by the fitting significance, $S_{fit} = W_{obs} /
\sigma_{W_{obs}}$, which is simply the ratio of the fitted equivalent
width and the uncertainty in that measurement.  For a more detailed
discussion, see \citet{petry06a}.  Since it is possible for a line
to be clearly detected but fit poorly (e.g., a blended line) and a
feature generated by noise to be reasonably fit with a Gaussian, these
significance parameters are used in tandem to assess absorption line
reliability.  Lines with both $S_{det} < 5$ and $S_{fit} < 2$ were
excluded from the Ly$\alpha$ sample.  These lower limits were selected
as a compromise between contamination by spurious lines and exclusion
of legitimate ones.  In a direct comparison of line identifications
for moderate resolution spectra and echelle spectroscopy of the same
targets, false positive and genuine line loss yields were found to be
approximately 8\% and 6\%, respectively (Marble \& Impey, in
preparation).  Due to high SNR, there are very few low significance
lines in our sample; only one line at 5114\AA\ in spectrum B (which is
blended with the wing of another line) was rejected using these
criteria (Figure \ref{sigs_f}).

\subsection{\lya Absorber Sample}\label{lyasample_ss}

As a prelude to analysis, we must ensure that the sample of absorbers represents only \lya 
lines in the IGM.  Damped systems appear in both spectra just blueward of \lya emission.  
The redshifts of these features are measured as $z = 3.778 \pm 0.003$ for A and $3.818 \pm 
0.003$ for B.  We performed a search for metal lines associated with these systems which 
produced statistically insignificant results.  To declare a match, the wavelength of a 
``matched'' line had to be within $2\sigma_{\lambda}$ of the redshifted metal line, 
corresponding to the redshift of the DLA.  With this procedure, five lines were matched to 
the DLA in A (\lyb, \lyc, \lyd, N II, Si III), and eight lines were matched to the DLA in B 
(\lyb, \lyc, \lyd, \lye, \lyf, \lyg, \lyh, C II).  Since the line density is so high at high 
redshift, we tested the significance of our matches by performing an offset experiment in 
which we shifted the spectrum and matched metals at each iteration.  This was done 400 times, 
from a 100\ang offset redward to 100\ang offset blueward using a 0.5\ang stepsize (the stepsize 
is larger than the wavelength error for a typical line).  The average number of random matches 
was 2.1 $\pm$ 1.4 in A and 3.2 $\pm$ 1.7 in B.  Since the number of metal matches were within 
$\sim2\sigma$ of the average number of random matches, we do not claim the identifications to 
be significant.  However, we contend that we can identify the higher order Lyman lines in A and 
B up to \lyd with good reliability.  The same offset experiment performed for higher order Lyman 
lines reveals a $2\sigma$ statistical excess at zero offset and the strengths of the lines in 
rough agreement with a predicted line strength.  These lines are removed from the \lya sample 
and marked as such in the Tables.

To determine the impact of the proximity effect \citep{scott00a} on absorbers close to 
\lya emission, we considered the luminosity of the quasars.  Both A and B are relatively faint, 
with apparent magnitudes $B_{A} = 20.0\pm0.03$ and $B_{B} = 20.7\pm0.03$ at redshift $\sim$3.8, 
suggesting the extent of the proximity effect is minimal.  The intervening DLAs further limit 
the extent of the proximity effect, to about 1000\kms, which does not include any lines in our 
lists (save the DLAs themselves).  Another consideration is the damped systems that in each case 
sit close ($\sim$800 \kms) to the emission redshift.  With a high column density, these features 
further limit the extent of the proposed Str{\"o}mgen spheres.  These coupled effects lead us 
to keep all lines fitted blueward of the DLA systems at $z = 3.778$ and $z = 3.818$, which is 
out to $\sim$1700 \kms.

\subsection{The \lyb ``Forest''}\label{lybforest_ss}

To test for \lyb contamination blueward of the \lyb emission line we performed an offset experiment.
The baseline, or zero offset, measurement involved assuming each absorber between the wavelengths of \lya and \lyb
emission is a \lya absorber, and predicting the wavelength positions of \lyb from each \lya line.  Then we shifted 
the wavelength centers of lines (blueward of \lyb emission) by offset increments of 5\angns, and measured the number 
of \lya$-$\lyb pairs.  The offsets ranged over $\pm$ 200\ang.  The criterion of accepting a pairing had to be that 
the observed line wavelength was within 2 $\sigma_{\lambda}$ of its predicted wavelength,
which was on the order of 0.4\ang (but determined uniquely from the corresponding \lya line).  Over the entire range 
of offsets, the average number of matched pairs was 8.5 $\pm$ 2.6 for A and 15.4 $\pm$ 4.0 for B.  At zero wavelength 
offset, the data show 10 pairs in A and 16 pairs in B, so there is no statistical detection of \lya$-$\lyb pairs. 

\begin{figure}
    \includegraphics[width=0.99\columnwidth]{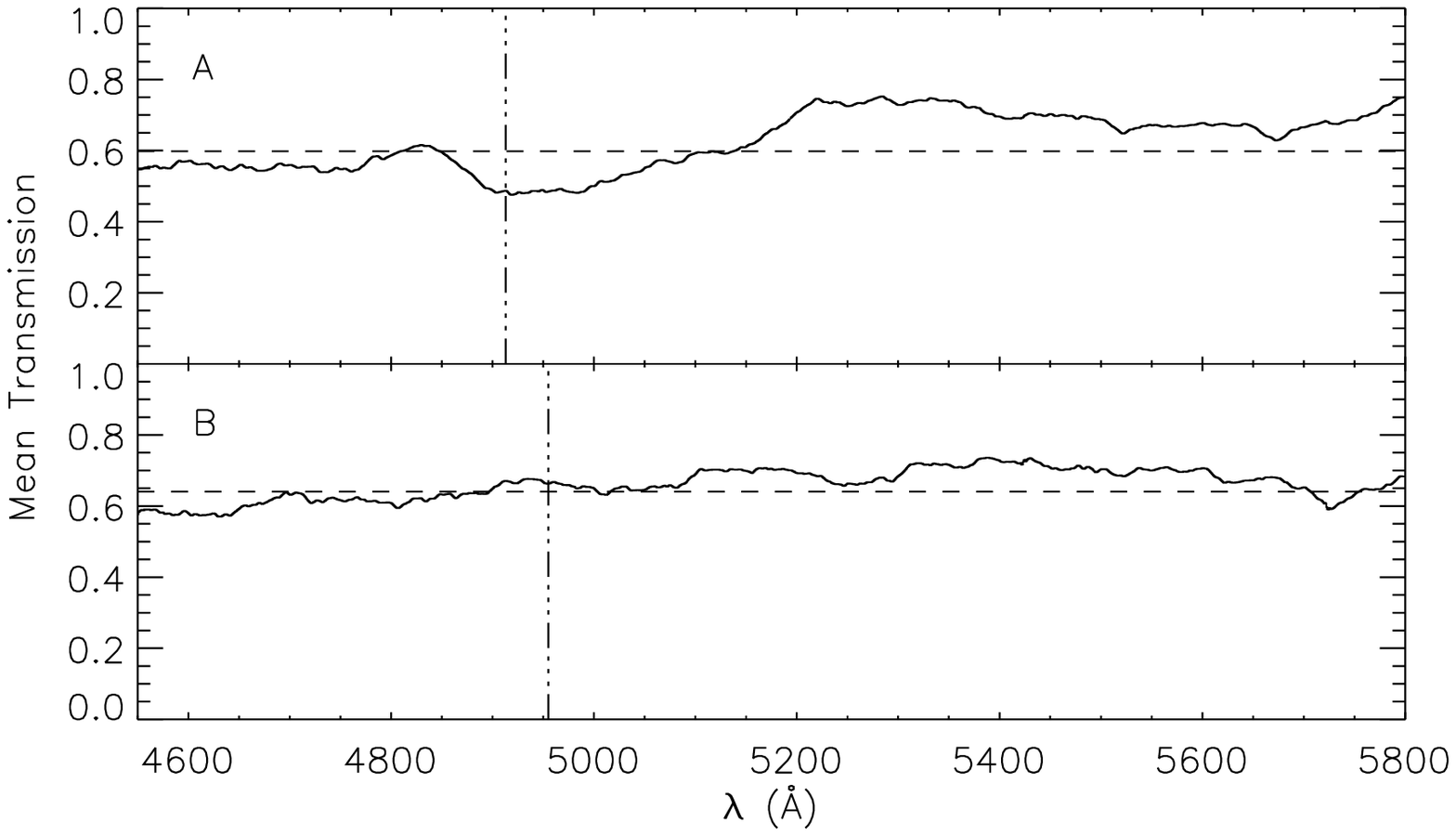}
    \includegraphics[width=1.10\columnwidth]{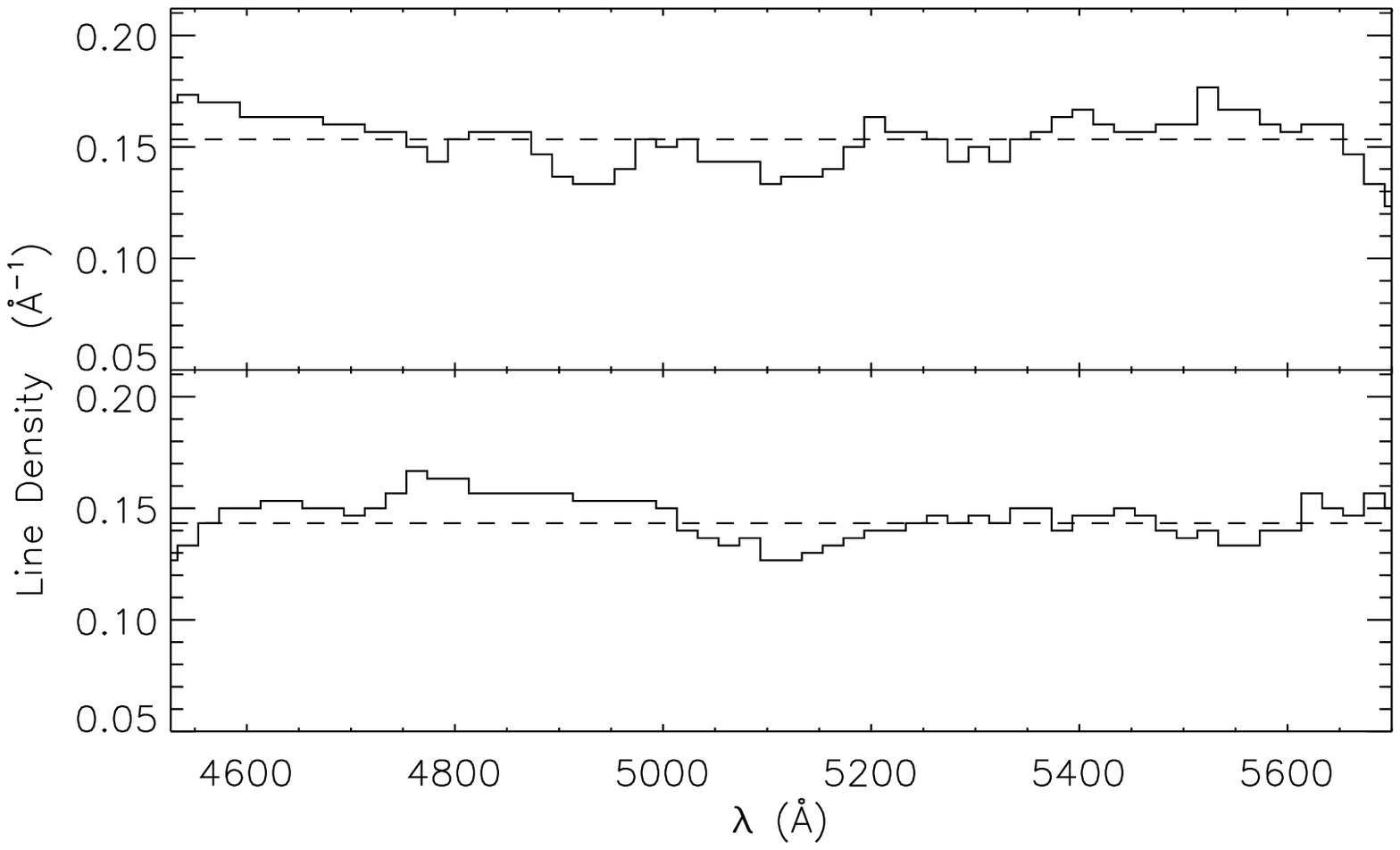}
  \caption{
    Mean transmission (averaged over $\pm$ 150\angns) vs. wavelength, showing little change 
    blueward of \lyb emission relative to redward (upper panel).  The vertical lines mark 
    \lyb emission for each quasar.  The lower panel shows the rolling average of the line 
    density in a 150\ang bin which is independent of equivalent width (since the distribution 
    in equivalent width does not significantly evolve with redshift).  Since we do not detect 
    extra absorption from metals or higher order Lyman lines and only observe slight changes 
    in transmission and line density, we carry forward into the analysis both (a) the entire 
    region from the Lyman limit to \lya emission and (b) the \lya forest alone.
  }
  \label{lyalyb_f}
\end{figure}

As further tests of whether we could use the full region down to the Lyman limit ($z \sim 2.65$)
to define a \lya absorber sample we looked at mean transmission and line density.  Figure 
\ref{lyalyb_f} shows that there is no significant change in mean transmission when passing from 
the pure \lya forest to the region where \lyb lines are also present, the ``\lyb forest.''  The 
line density (calculated independent of equivalent width) is also shown in Figure \ref{lyalyb_f} 
and is constant across the entire range of data, supporting the assumption that most lines are 
\lya unless otherwise noted.  This line density comparison is valid since the observed distribution 
of equivalent widths does not change as a function of redshift.  Our discrete absorber analysis
includes the \lyb region, while our flux statistics studies and cross-correlation calculations
proceed through analysis with and without exclusion of the \lyb forest, which will be discussed more 
thoroughly in \S \ref{sph_s}.

\subsection{Statistical Absorber Properties}\label{dnewdz_ss}

\begin{figure}
  \includegraphics[width=0.99\columnwidth]{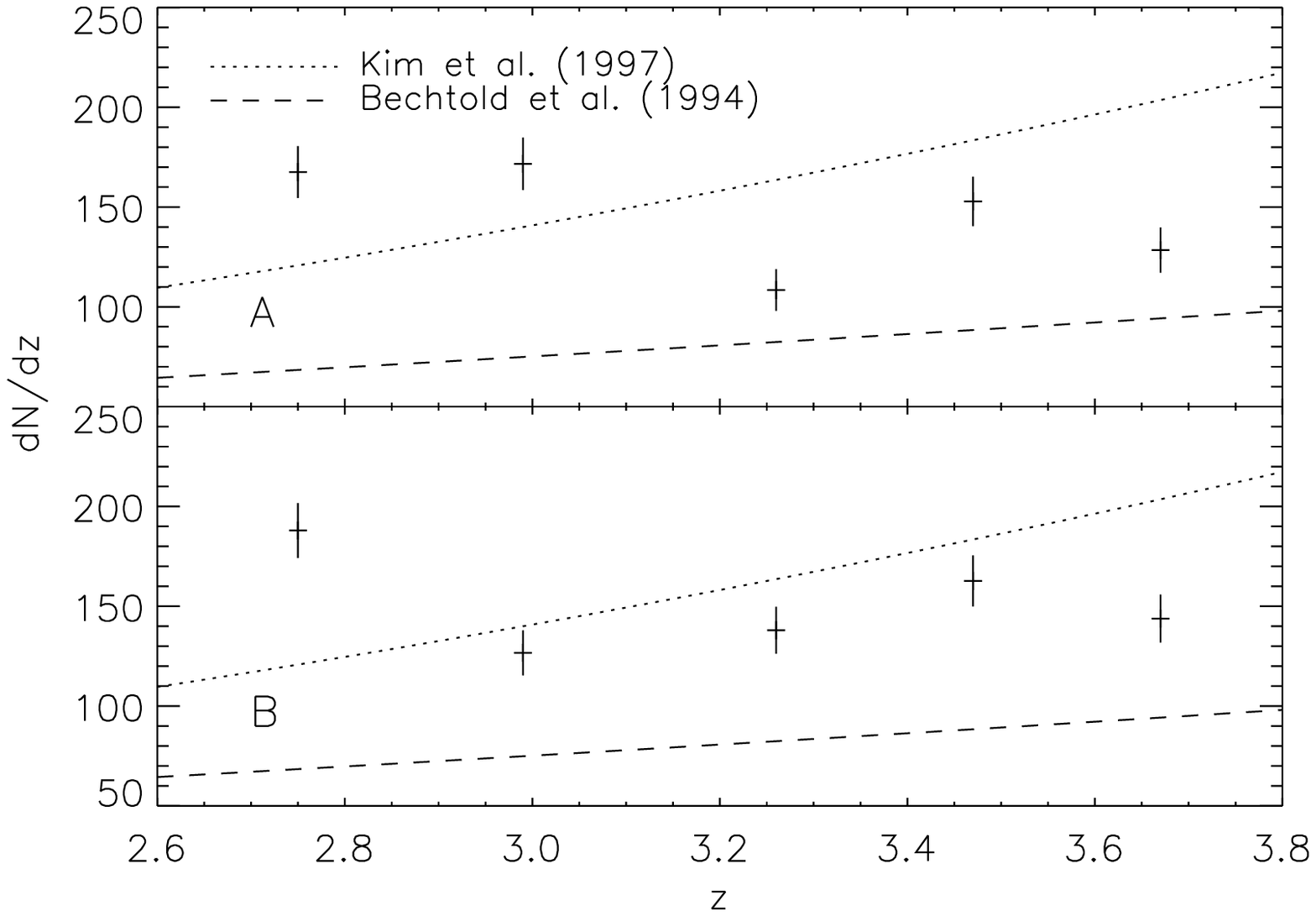}
  \caption{
    \lya absorber density as a function of redshift for lines with $W_{0} > 0.32$\angns, compared to two previoius studies
    \citet{bechtold94a}, and \citet{kim97a}, both with lower limits on equivalent width of 0.32\angns.  The higher line density
    of the latter study is due to better signal-to-noise and resolution.
    Line density is calculated in five bins (avoiding the strong DLA in spectrum A at $\sim$5030\angns) centered
    on redshifts 2.75, 2.99, 3.26, 3.47 and 3.67.  No evolution of strong lines is apparent since we underestimate 
    the continuum and thus underestimate line counts and line strengths at higher redshifts.
  }
  \label{dndz_f}
\end{figure}

The line culling process described in the preceding sections (removing
lines from metal contamination, proximity effect, poor fitted or detection
significance, or damped system lines), reduces the full line lists in
Tables \ref{tab1} and \ref{tab2} to the \lya sample: 222 lines in A and
211 lines in B. In accordance with previous studies, we express the number
of absorption lines per unit redshift per unit rest equivalent width as
\begin{equation}
  \frac{\partial^{2} N}{\partial z \partial W} = \frac{A_{0}}{W*} (1 +
z)^{\gamma} exp\{-\frac{W}{W*}\}
\end{equation}
where $\gamma$ and $W*$ are determined by maximum likelihood estimation as
in \citet{murdoch86a} with code written by A. Dobrzycki.  We can also
express the number of lines per unit redshift above a fixed equivalent
width limit (often taken in the literature as 0.32\ang or 0.16\ang), as
\begin{equation}
  \frac{dN}{dz} = A_{0} (1 + z)^{\gamma}
\end{equation}
The most appropriate line density comparison from the literature
comes from \citet{bechtold94a}, who presented a moderate resolution, high
redshift ($z > 2.6$) sample (their Table 4, sample 21a for $W_{thr} =0.32$\ang 
and sample 22a for $W_{thr} = 0.16$\angns), and from
\citet{kim97a} (likewise from \citealt{hu95a}), who used high resolution
data at all redshifts (with $W_{thr} = 0.32$\angns).  We adopt the
equivalent width threshold value of 0.32\ang since it is above our
limiting equivalent widths for A and B at nearly all wavelengths.
Splitting the data into five redshift bins each centered at redshifts
2.75, 2.99, 3.26, 3.47 and 3.67 (excluding data from the damped feature in
A surrounding 5005\angns-5060\angns) we present $dN/dz$ in Figure
\ref{dndz_f}.  The difference between the two earlier studies is readily
understood as an effect of line blending and limited resolution. Kim et 
al. did simulations to show that their completeness to 0.32\ang lines
at $z\sim 3$ was about 90\%.  Bechtold only was able to recover about
half of the lines at this strength, and our data is intermediate in
completeness.

We see broad agreement with the two earlier studies except
in the lowest one (for B) or two (for A) redshift bins.  Since the \lyb
forest begins in the middle of the second lowest redshift bin, around $z\sim 3$, 
this casts some doubt on using this region in the \lya analysis.
From this direct comparison of $dN/dz$ with literature values of high
resolution data, we estimate the number of absorbers that are averaged
together due to the spectral resolution and SNR.  As a function of
redshift (split into the five redshift bins earlier described at z=2.75,
2.99, 3.26, 3.47, and 3.67), the number of hypothetical \lya absorbers
divided by the observed number of absorbers is 0.67, 0.96, 1.36, 1.51, and
1.50 (i.e. blending of $\sim 3$ true \lya absorbers into $\sim 2$ fitted
lines at the highest redshifts).

While the additional signal to noise in spectrum A might lead one to believe that more absorbers
should be fit than in spectrum B, the resolution and high line density of this high redshift data 
washes out the dependence of line count on signal to noise directly.  We fit all lines with the 
assumption that they are unresolved, and had line profiles equal to the resolution of our 
observations (3.6\angns).  The separation between adjacent fitted lines is on order 2-4 times 
this wavelength interval.  Since the separation between lines and width of lines is comparable, 
and lines are restricted to have a minimal seperation of at least one sampling unit (0.85\angns), 
it can have a enormous effect on the effective line density of the data.  Weaker features are 
washed out of the experiment due to the resolution limit of line fitting.  So the additional lines 
that one might expect to exist in spectrum A were likely neglected due to blending.  When fitting
lines to the simulation spectra (as part of our comparison with simulations in \S \ref{simlines_ss}), 
we find that the number of lines fit to either sightline (A or B) are comperable, and S/N does not
have an effect on the lists from resolution limitations and a very high line density.

\begin{figure}
  \includegraphics[width=0.99\columnwidth]{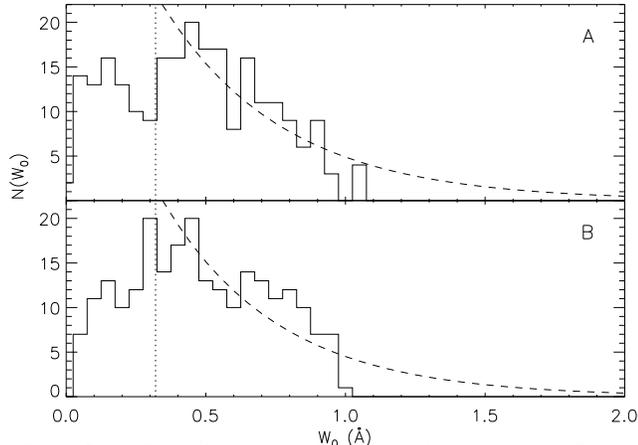}
  \caption{
    The distribution in rest equivalent widths for \lya absorbers from both sightlines.  
    Above 0.32\angns, the distributions are moderately well-described by exponentials.  
    The exponential fits (dashed lines) are described by $N_{A}(W_{0}) = 48.42$exp$(-2.29W_{0})$ 
    for A and $N_{B}(W_{0}) = 50.91$exp$(-2.42W_{0})$ for B.  The distribution shapes 
    are similar, and when split into higher and lower redshift bins show no sign of evolution.
  }
  \label{dndw_f}
\end{figure}

The distribution in rest EW for both \lya forests is shown in Figure
\ref{dndw_f} with exponentials overplotted that only fit data above the
$W_{thr} = 0.32$\ang limit.  Above this limit, the distribution is
well-described by an exponential.  At high redshift ($z > 3.4$) the
continuum fit from \S \ref{contfit_ss} is suspected to significantly
underestimate the true continuum level (discussed quantitatively in \S
\ref{underestimation_ss}), which means that the equivalent widths of high
redshift absorbers are underestimated.  We observe no evolution of the
distribution of equivalent widths with redshift.  The effect on line count
and line strength from continuum underestimation is complex, based both on
the omission of weaker absorbers and the underestimation of equivalent
widths from stronger features.  This systematic bias in the continuum
applies equally to both sightlines and does not affect measures of the
transverse coherence between discrete absorbers (since the effect is equal
for both sightlines), but it will significantly impact measures of
transverse correlation when considered in flux statistics tests, which is
why we quantify the underestimation in \S \ref{underestimation_ss}.

\section{ABSORBER COINCIDENCES}\label{linecoincide_s}

\subsection{Symmetric Matching}\label{symmmatch_ss}

\input{tab3}

Symmetric pairing of lines across the spectra is done via nearest neighbor matching in wavelength space.  
A pair is considered symmetric if both lines are each others' nearest neighbor, 
i.e. the line from A is the nearest feature in wavelength to the line in B and vice versa.  This 
matching method is the simplest way to measure coherence with
discrete measures of opacity like absorption lines.
We find 152 matches across the sightlines; the specifics of 
the absorber properties and separations are listed in Table \ref{tab3}.  Next
we explore the properties of the symmetric matches in this dataset, the scale on which we expect 
matches to be made, and whether or not coherence can be inferred from this high redshift data relative to Monte
Carlo simulations.

\subsection{Characteristic Absorber Separations}\label{charlength_ss}

\begin{figure*}
  \centering
  \includegraphics[width=1.90\columnwidth]{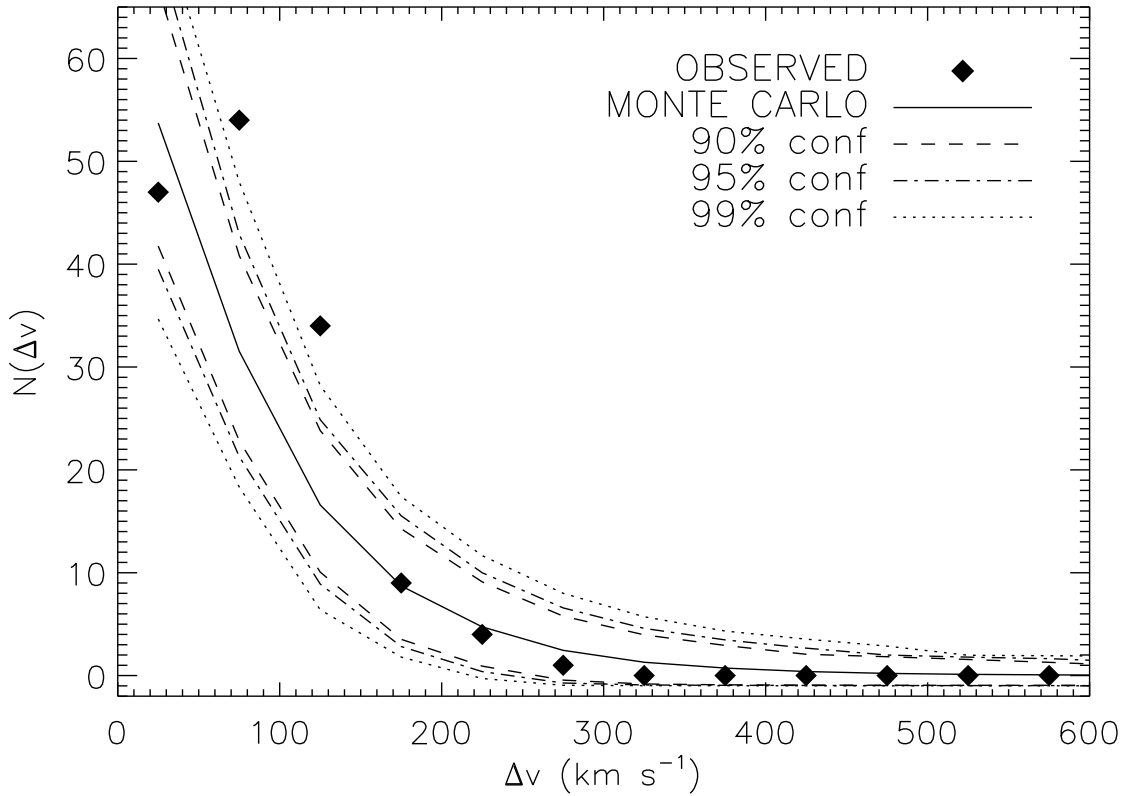}
  \caption{
    The velocity splitting of 152 symmetrically matched pairs.  The results of random matching 
    Monte Carlo experiments are overplotted with 90$\%$, 95$\%$, and 99$\%$ confidence intervals.  
    The first data bin, from 0-50 \kms, is significantly limited by the instrumental resolution 
    and thus contains fewer matches than high resolution echelle spectra would exhibit.  The 50-100 
    \kms\ and 100-150 \kms\ bins show a 3-4$\sigma$ coherence signal above the Monte Carlo expectation.   
    Wider velocity splitting bins with $\Delta v > 150$ \kms\ follow expectation from random pairings.
  }
  \label{montecarlo_f}
\end{figure*}

Symmetric pair matches may just be chance pairings.  The 
mean separation between matched pairs is 1.4\ang or $\sim$84 \kms, as seen as a peak in Figure \ref{montecarlo_f}.  
In contrast, the average separation between adjacent lines in each individual spectrum 
(a measure of line density) is 6.1 $\pm$ 3.6\ang ($\sim$370 \kms) in A and 6.8 $\pm$ 4.2\ang ($\sim$410 \kms) in B.  
If the requirement of symmetry (A matches to B and B matches to A) is dropped in the matching process, the 
mean $\Delta\lambda$ increases as expected (to 3.4$\pm$6.5\angns).
To indicate the physical scale on which lines are being matched, 
we calculate the velocity scale of differential Hubble flow.  Using an angular separation
of 198\arcsec, a typical radial line separation translates to $\Delta\lambda_{o}$ =
4.5\angns, or $\sim$270 \kms.  

\subsection{Monte Carlo Experiment}\label{montecarlo_ss}

We perform a Monte Carlo experiment to test the significance of symmetric line matches, by 
making matches across sightlines as a way to detect coherence (above the chance matching 
resulting from randomly placed lines).  This test is not limited or compromised by low 
resolution because lines are typically centroided to 30 \kms ($\sim$0.5\angns), which is 
$\sim$7 times smaller than our 220 \kms (3.6\angns) resolution \citep[][, using the same 
method for HST data of similar resolution]{dinshaw98a}.  Since the line velocity error 
(30 \kms) is much less than the mean velocity splitting of matched pairs (84 \kms), the 
measurement error does not impact this line matching coherence measure. With 1000 
realizations, the experiment recreates the two sightlines, sampled to mimic the line 
density and equivalent width distribution of the observed spectra, and computes symmetric 
matches between the two randomly drawn sets of absorbers.  The entire range of the data is 
used, including the \lyb forest.  This adds needed line statistics to the \lya sample, and 
since line fitting is somewhat immune from varying mean flux and continuum fits (which 
exhibit strange behavior in the \lyb region as can be seen by excessive opacity in Figure 
\ref{allspec_f}, particularly in A) there is no concern that line matches in the \lyb region 
are not meaningful.  The only risk in including the \lyb region is double counting \lya and 
their \lyb counterpart matches, which will also trace coherence in the IGM.  This double 
counting is unlikely since we found very little evidence that \lyb contamination is significant, 
and at most, would only change the result by $\sim$10$\%$.

To create the simulated Monte Carlo samples, we use the limiting equivalent width (or detection limit) 
as a function of wavelength, which hovers about 0.24\angns.  Damped regions are removed from 
the simulated spectra as from the data: 5005-5060\ang in A and 5145-5160\ang in B.  We overplot the 
results of this Monte Carlo matching experiment on our observed matches in Figure \ref{montecarlo_f}.
The data show a $\sim$4$\sigma$ coherence signal above the Monte Carlo random 
matching result in the 50-100 \kms\ and 100-150 \kms\ bins (pair excesses are 3.9$\sigma$ and 3.7$\sigma$ 
respectively).  The experiment was repeated by splitting the sample into strong and weak lines (split at 
0.5\angns, roughly half the sample), but it did not show that coherence was more prevelent in either strong 
or weak absorbers.  The first bin (0-50\kms) is depopulated due to line blending; 
few symmetric matches on small velocity splitting scales would be counted if blended features containing 
two absorbers correspond to single absorbers in the opposite sightline.  Therefore, we do not claim that the 
offset signal (peak between 50 \kms\ $< \Delta v <$ 150 \kms) is due to inherent shear in the IGM.

\section{COMPARISON WITH THE SIMULATIONS}\label{sph_s}

\subsection{Absorber Coincidences from Simulations}\label{simlines_ss}

\begin{figure}
  \includegraphics[width=0.99\columnwidth]{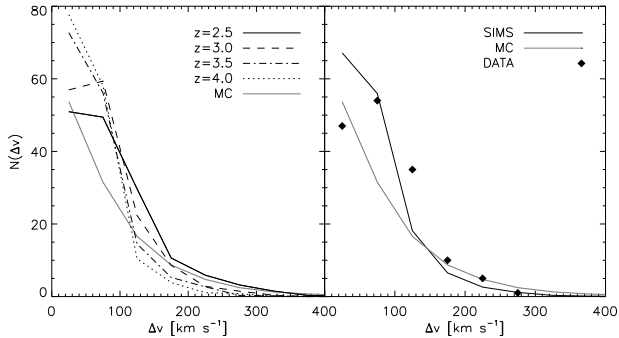}
  \caption{
    The velocity splitting distribution of absorbers in the spectra
    extracted from the simulations.  The left panel shows the
    distribution as a function of realization redshift.
    For reference, we have overplotted the random
    absorber result from Monte Carlo tests (as shown in Figure \ref
    {montecarlo_f}), and we see that the simulations show a weak but
    detectable coherence signal above random chance pairings.  When
    properly weighted over the redshift interval of our data (2.6 $<$ z $<$
    3.8), we show the simulations alongside the data result and the
    random test result in the right panel.
  }
  \label{sims_vs_f}
\end{figure}

To further interpret our $\sim$4$\sigma$ detection of coherence from
the previous section, we perform a similar symmetric pair matching on
absorbers in the spectra extracted from simulations (as described in
\S \ref{g6sims_s}).  Line fitting was done with the same $ANIMALS$
utility that was used for the data, this time applied to the degraded
extractions from simulations designed to match the data.  Figure \ref
{sims_vs_f} (left panel) shows the velocity splitting distribution at
each redshift for symmetrically matched pairs in associated A and B
simulated spectra (with the MC randomization result overplotted for
reference).  At higher redshift, greater line densities generate a
steeper peak at zero splitting as expected, while lower redshifts give
shallower distributions.  To make a fair comparison with the data, we
interpolate the distribution shape that simulations would produce for
a spectrum continuously sampling redshifts 2.6 $<$ z $<$ 3.8.  This is
shown alongside the randomized result and the data in the right panel
of Figure \ref{sims_vs_f}.  This not only reaffirms that we can
detect weak coherence in the lya forest at high redshifts, but that
the simulations agree with our data in showing a signal of roughly
the same strength.

The one area of disagreement between data and degraded extractions
from the simulations is in the first velocity offset bin (0-50\kms).
This can be explained simply as a depopulation of the smallest velocity
offset bin in the data due to line blending (i.e. the discrepancy is
non-physical).  Although this happens to some degree in the spectra
extracted from simulations, the settings in $ANIMALS$ used to generate
line lists from the simulations allow lines to be fitted with smaller
wavelength separations (below the 0.85\ang restriction applied to the
data) since the spectra themselves are so short (only 300\ang,
rather than the data that span $\sim$1500\angns).  The lowest bin,
where the turn-down occurs in the data, correponds to splittings less
than 40$\%$ of the instrumental resultion.  At these high redshifts, where
blending is severe, the transverse matching result will be sensitive to
the exact criterion used for minimum line separation in each sightline.
The slight depopulation of simulation matched pairs relative to data
at high velocity splitting is a consequence of the same effect.  To
first order, this preliminary test for high-z coherence is successful
and it encourages additional testing of the simulations that can be
done in the future more effectively using higher quality data.

\subsection{Evolution in the Transmission Distribution}\label{transhist_ss}

\begin{figure}
  \includegraphics[width=0.99\columnwidth]{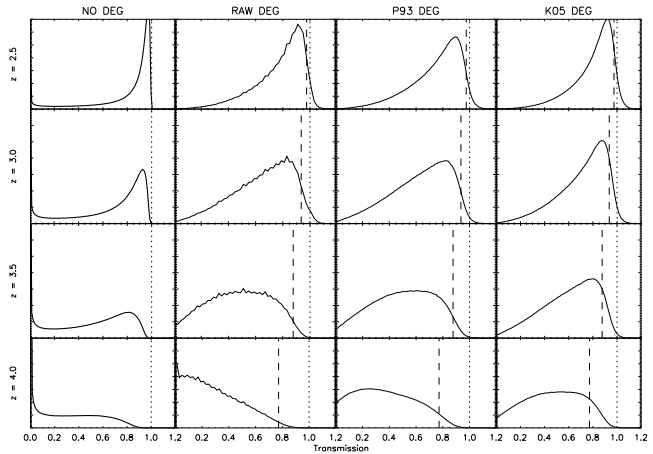}
  \caption{
    The transmission distributions of 3000 spectra extracted from SPH simulations at four discrete redshifts.  The first column
    shows the distributions of non-degraded simulation spectra, with no noise added and very high resolution, which 
    characterizes the intrinsic cosmic evolution of the IGM in the simulations.  The second column
    shows the distributions for simulations after degrading to the signal-to-noise and instrumental resolution
    of the spectra, but without rescaling the mean flux.  The distributions of A and B 
    do not differ despite a difference in SNR, so they are combined.
    The effect of resolution is particularly visible at high redshift.  The third and fourth
    columns represent the same distributions from column 2 after being rescaled by mean flux according to P93 \citep*{press93a}
    and K05 \citep{kirkman05a} respectively. The vertical lines in the three last columns represent the average 
    transmission value of the fitted continua as described in \S \ref{underestimation_ss}.
  }
  \label{transhist_f}
\end{figure}

The spectra extracted from the G6 simulation (\S \ref{g6sims_s}) show significant changes 
in opacity distributions over the redshift range of our data.  At their initial high 
resolution and infinite SNR, the transmission distributions show substantial absorption 
at higher redshift while very little at low redshift, the anticipated strong cosmic evolution 
from z = 2.6 to z = 3.8 (Lyman limit to \lya emission in our data).  The 
transmission distributions in the first column of Figure \ref{transhist_f} 
combine the statistics of both A and B simulated sightlines, non-degraded 
and not rescaled to mean flux via either P93 or K05. At z = 2.5, the lowest 
redshift of simulation extractions, there is a strong peak at high transmission 
indicating little absorption relative to the true continua; the small tail 
at zero transmission indicates the number of pixels in damped systems.  At 
higher redshifts, z = 3.0 and 3.5, there is a clear shift in the distribution 
towards lower transmission, and the z = 4.0 distribution suggests opacity at 
nearly every pixel.  The number of opaque pixels increases tenfold from z =
2.5 to z = 4.0 from 3.1$\%$ to 20.0$\%$ of all pixels. 

After degrading the raw extracted spectra to the signal-to-noise and resolution of the 
data, the second column of Figure \ref{transhist_f} shows changes in the overall transmission 
distribution at each redshift. There was no observed SNR effect on transmission distributions 
shape, so the statistics from both fabricated sightlines were combined and weighted by relative 
SNR.  The effects are as expected: the peaks have broadened, there are fewer pixels with zero 
transmission, and since noise has been added, a small fraction of pixels have transmission 
greater than one.  With increasing redshift, the distribution shape evolves from a high 
transmission peak to a nearly linear fall off with higher transmission at z = 4.0.  In the 
figure, the dotted line indicates zero opacity (the true continuum), and the dashed line 
indicates the average level of the continuum fit after the simulation extractions are degraded.

If the degraded simulations are altered to match the mean flux estimates given 
by P93 and K05 the transmission distributions change shape, as shown in the third and fourth columns of Figure 
\ref{transhist_f}.  The P93 rescaled simulations have very similar shapes to the raw degraded 
simulations, while the K05 rescaled simulations do not show as much dramatic evolution at high redshift.
This comparison shows the importance of mean transmission in interpreting data where it is not known a priori.

\subsection{Underestimation in Continuum Flux}\label{underestimation_ss}

From the simulation, we infer a systematic error in $ANIMALS$ continuum fitting as a 
function of redshift.  When degraded to the resolution of our observations, the 
smoothing of transmission on small velocity scales shifts the peak transmission 
values lower, and so depresses any algorithmic fit to the continuum.  This effect 
becomes larger at higher redshifts as the mean opacity in the IGM increases.  The 
vertical dashed lines on columns 2-4 of Figure \ref{transhist_f} represent the
transmission values of the fitted continuum (with respect to the true continuum level).
When we compare the data's transmission distributions to those of the simulation, we
use the estimate of the continuum fit transmission to rescale the data's flux, thus 
correcting the underestimation and making the comparison between data and simulation fair.
The flux decrement of the continuum may be modeled as a parabolic function of $(1+z)$:
\begin{equation}\label{contunder_eq}
  fd_{cont}(1 + z) = 0.697 - 0.419 (1 + z) + 0.065 (1 + z)^{2}
\end{equation}
with variance $\sigma_{fd}^{2} = 0.002$, and valid for $2.5 < z < 4.0$.  A quadratic fit 
is empirically chosen because a linear fit is an inappropriate choice to model such redshift 
evolution.  The rescaling of degraded spectra to either P93 or K05 does 
not have any additional impact on the underestimation of continuum flux.  Pixels that have
very high or low transmission are not affected by rescaling the mean flux as much as pixels with 
intermediate absorption, and since the continuum fitting is based on the highest transmission
pixels, the rescaling effect will be minimal, particularly at low redshift.  At high redshift
the size and uncertainty in the continuum underestimation grow as would the effect
of rescaling by mean flux, but it is clear that the systematic error in the fit dominates
the continuum flux value over the smaller variations between raw, P93, and K05 mean flux rescalings.

This systematic effect leads to an underestimation of line density
as well as equivalent width.  At high redshift, where the effect is most severe, only the
strongest and highest column density features will be fit via $ANIMALS$ methods, while the much weaker
lines in blended regions are not fit.  The size of this effect cannot 
be properly measured since line fitting relative to the true continua is 
not a unique process when there is opacity at nearly every pixel.  Echelle spectra
can mitigate this problem, since moderately strong features will be fully resolved and reliably
measured regardless of the true continuum level.

\subsection{Comparisons of Transmission Distributions}\label{datavsims_ss}

When comparing the observational data to simulations we must recall that the SPH 
extractions represent discrete epochs (z = 2.5, 3.0, 3.5, and 4.0) while the data 
samples redshift continuously with cosmic evolution superimposed.  We split the 
spectra into five bins $\sim$300\ang wide, centered at redshifts 2.75, 2.99, 3.26, 
3.47 and 3.67.  These redshift bins were chosen to avoid problem areas in both 
spectra$-$the large damped system in quasar A (5005-5060\angns), all data blueward 
of the Lyman limit in either spectrum ($\lambda > $4420\angns), and all data redward 
of the DLAs close to the \lya emission peaks ($\lambda < $5790\angns).  The first two 
bins, z = 2.75 and 2.99, cover the \lyb forest while the three subsequent bins, z = 
3.26, 3.47 and 3.67, cover the \lya forest and are slightly narrower in width 
($\sim280$\angns).

\begin{figure}
  \includegraphics[width=0.99\columnwidth]{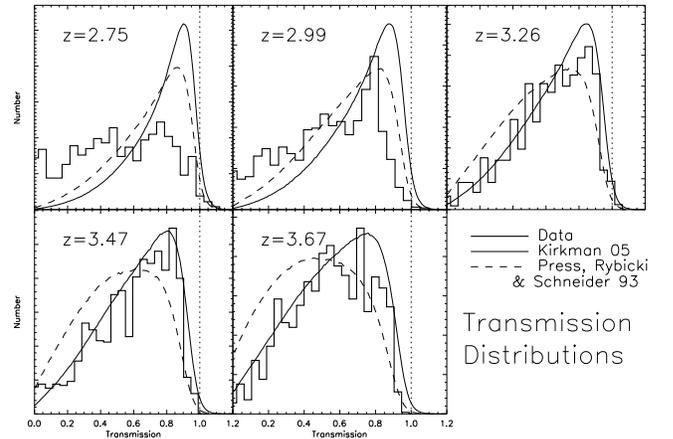}
  \caption{
    Transmission distributions of combined data for both quasars split into five redshift 
    bins centered at 2.75, 2.99, 3.26, 3.47, and 3.67.  The distributions of spectra 
    extracted from simulations are overplotted at these redshifts by linearly interpolating 
    the transmission distribution shapes from the four extraction redshifts: 2.5, 3.0, 3.5, 
    and 4.0.  \citet*{press93a} (dashed lines) and \citet{kirkman05a} (solid lines) mean flux
    rescalings show quite different distribution shapes, particularly at high redshift. The 
    data transmission was rescaled to correct for the systematic underestimation in the 
    $ANIMALS$ continuum fit, as described in \S \ref{underestimation_ss}.
  }
  \label{datasims_f}
\end{figure}

We linearly interpolate the shapes of the simulations' transmission distributions 
to infer the distribution shape at these five intermediate redshifts.  We do these 
procedures using both P93 and K05 sets of spectra, adopting different mean flux 
values for the simulations.  Figure \ref{datasims_f} shows the transmission 
distributions for the data (histogram) and the simulations (P93 scaling is dashed 
and K05 scaling is solid) at the five redshift bins.  Since the continuum flux has 
a significant effect on transmission (particularly at high redshift), we rescale 
the normalized data flux by a factor representing the mean $ANIMALS$ fitted continuum 
flux (as a function of redshift).  The transmission of each pixel in the data is 
rescaled by this continuum flux factor, given by equation \ref{contunder_eq}, which 
removes the effect of the systematic underestimation in the continuum fit. 

At the lowest redshift (z = 2.75 in Figure \ref{datasims_f}), the data show far 
more low transmission pixels than anticipated.  The extra absorption relates 
directly to the excess in line density seen in Figure \ref{dndz_f}, but is too 
large an effect to be attributed directly to \lyb absorbers or metal absorbers.
Since the signal-to-noise of the blue end of the data is quite low, the true 
transmission distribution may be modified by a large flux error.  The remaining 
four bins from 2.99 to 3.67 show fairly good agreement between simulations and 
data.  To test agreement with the simulations models, both from the Kirkman et 
al. and Press et al. treatments, we performed K-S tests to test goodness of fit.  
It revealed that significant agreement ($>$45$\%$) only occured in the three 
highest redshift bins (z = 3.26, 3.47 and 3.67) using the Kirkman mean flux 
scaling simulations treatment (with probabilities of 48$\%$, 96$\%$, and 63$\%$ 
respectively).  The lower redshift bins in the Kirkman treatment, and none of the 
\citet*{press93a} transmission distributions were proper fits, all with probabilities 
lower than 3$\%$.

\subsection{Transverse Correlation Comparison}\label{andystat_ss}

\begin{figure}
  \centering
  \includegraphics[width=0.99\columnwidth]{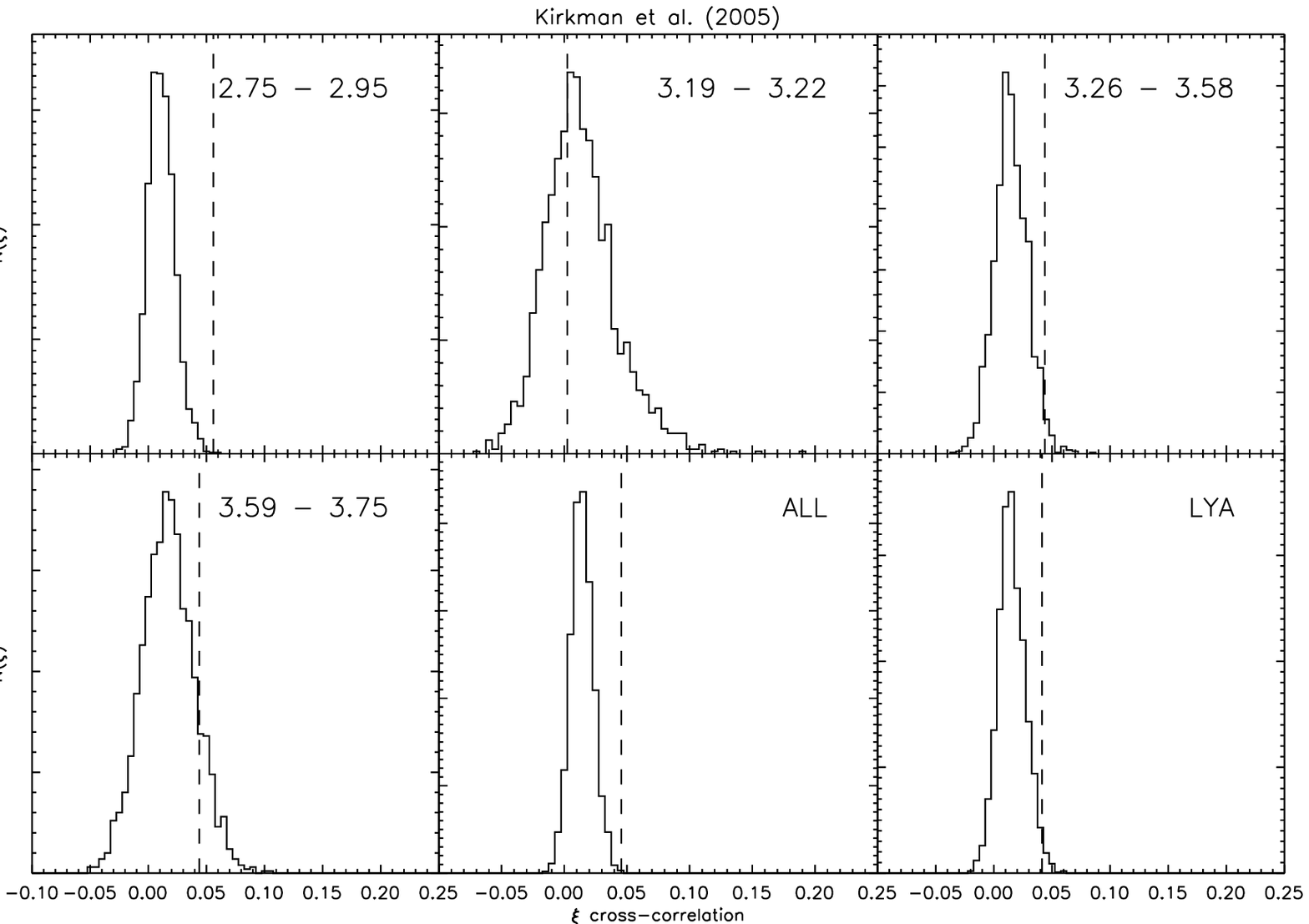}
  \includegraphics[width=0.99\columnwidth]{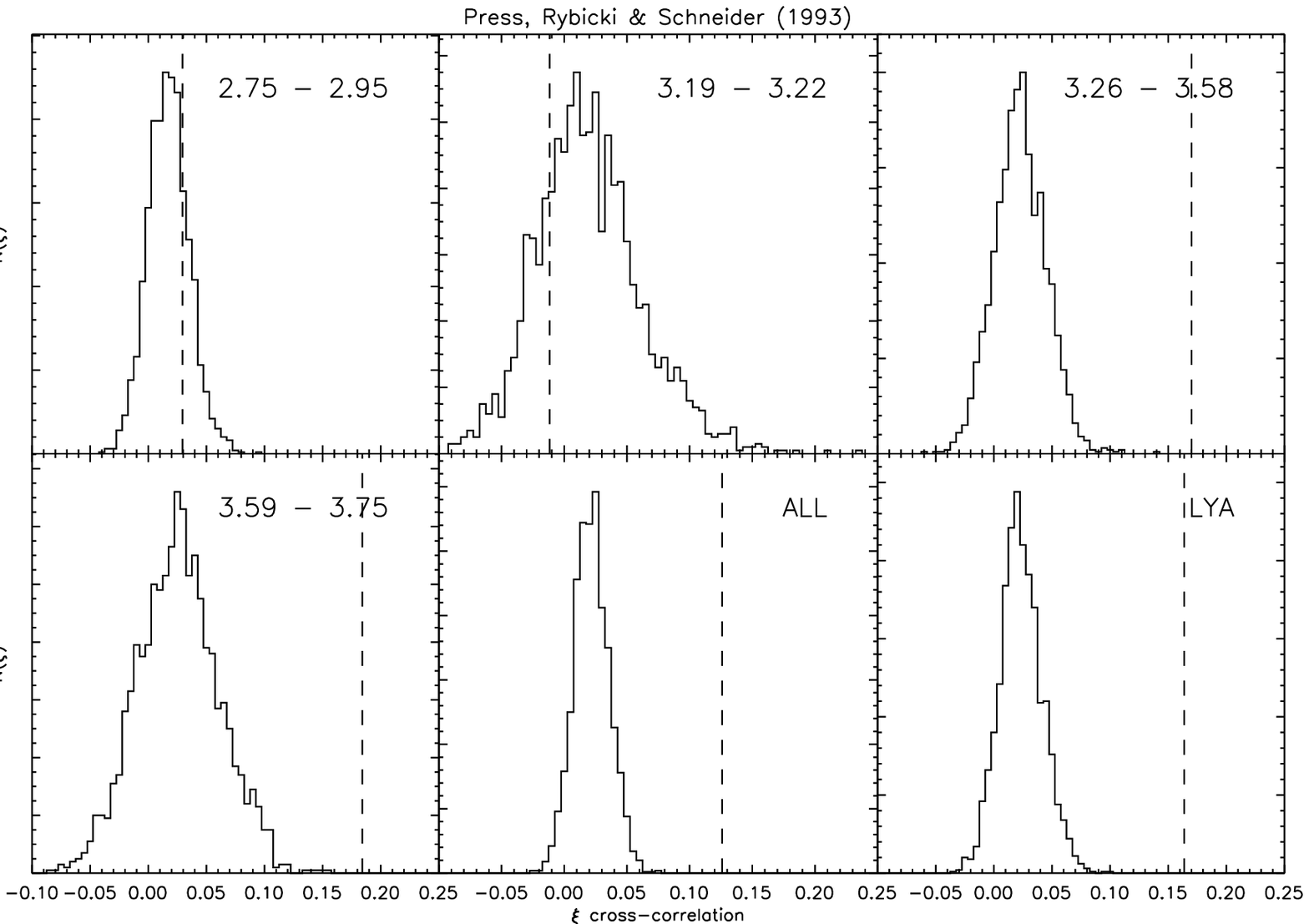}
  \caption{
    The distribution in cross-correlation amplitude for the simulations, with \citet{kirkman05a} 
    (top) and \citet*{press93a} (bottom) mean flux rescalings.  The 2000 simulations used for 
    this experiment are split into four (uneven) redshift bins to reproduce the cleanest portions 
    of observed spectra: $2.75 < z < 2.95$, $3.19 < z < 3.22$, $3.26 < z < 3.58$ and $3.59 < z < 
    3.75$.  The first bin shown includes a portion of the \lyb forest.  The final two panels are 
    combinations of the first four: 'ALL' means it covers the entire range of our data (all four 
    bins) and 'LYA' represents the exclusive \lya forest experiment, $z > 3.19$ (excluding the
    lowest redshift bin).  The dashed vertical lines indicate the cross-correlation measurements 
    for the data.
  }
  \label{simcrscor_f}
\end{figure}

To compute the cross-correlation between simulation sightlines, the extracted spectra 
are pieced together from the z = 3.0 and z = 3.5 realizations to cover the wavelength 
range of the data.  Instead of using linear interpolation to account for evolution at 
intermediate redshifts, piecing together extraction spectra (300\ang in length) from
the discrete redshift bins (2.5, 3.0, 3.5, 4.0) is a better test of simulation and data
agreement.  Rather than measuring cross-correlation for each redshift and then 
interpolating, we meaasure the cross-correlation from spectra pieced together to model 
the redshift evolution.  The spliced simulation spectra run from 4420\ang (the Lyman 
limit in B) at the blue end to 3000 \kms\ blueward of \lya emission in A (5765\angns).  
Portions of the spectra are masked due to features in the data: the damped system in A 
from 5005-5060\angns, an emission feature in A from 5142-5148\angns, and the 5575\ang 
sky line.  Since inclusion of the \lyb forest region may not be warranted (see the 
discussion in \S \ref{dnewdz_ss}), we perform cross-correlation in segments: from $2.75 
< z < 2.95$ (the \lyb region with $SNR > 10$), $3.19 < z < 3.22$ (the \lya forest between 
the damped feature and \lya emission), $3.26 < z < 3.58$ (the \lya forest between \lya 
emission and the 5577\ang sky line), and $3.59 < z < 3.75$.  The results are then combined 
into two larger bins: the \lya forest (all bins with $z > 3.19$) and the entire range 
(including the lowest bin, which spans part of the \lyb forest).  The cross-correlations 
are computed for both P93 and K05 mean flux scalings.  The distribution of the 
cross-correlation amplitudes for the simulations is shown in Figure \ref{simcrscor_f}.  
The top set of panels shows the results for \citet{kirkman05a} mean flux rescalings while 
the bottom half show those for \citet*{press93a}.  The cross-correlation amplitude is 
within 1-2$\sigma$ of zero in every case, which emphasizes the difficulty of measuring 
coherence at this physical separation and redshift, given sample variance.  Overall, the
experiment shows that the K05 rescaled simulations are correctly modeling the \lya absorbers 
at high redshifts since the data measurements are consistently within $\sim$1$\sigma$ of
the mean cross-correlation amplitude.  On the other hand, simulations assuming rescaled flux
according to \citet*{press93a} do not agree with data, with a difference of $\sim$8$\sigma$
over the \lya region.  This agreement with \citet{kirkman05a} rather than 
\citet*{press93a} is consistent with the earlier result from transmission distributions in 
\S \ref{datavsims_ss} (as seen in Figure \ref{datasims_f}).

\section{SUMMARY}\label{summary_s}

This paper has presented new spectroscopic observations of the $z \sim 3.8$, 198\arcsec\ separation 
quasar pair PC 1643+4631A, B and associated detection of coherence in the IGM on scales that have not 
been previously tested, $\sim$2.5 $h^{-1}_{70}$ Mpc.  The observations cover the full extent of the 
\lya forest range from \lya emission to the Lyman limit (with high signal-to-noise and moderate 
resolution, 3.6\angns) and provide an excellent opportunity to not only measure coherence and large 
scale structure in the IGM at high redshift, but also compare observations with predictions from 
cosmological simulations.

The \lya absorber sample was defined in two ways: using the entire range of data from \lya emission to the Lyman 
limit, or restricting the \lya experiment to the region  between \lya emission and \lyb emission, dubbed 
the pure \lya region.  Metal and higher order Lyman lines were not detected in the \lyb forest region 
due to the high line densities, but since contamination may be present and the blue wavelengths have low 
SNR we treat the \lyb forest region separately from the \lya forest in flux statistics studies.  A Monte 
Carlo random pairing experiment using the entire range revealed that chance pairs account for a significant 
portion of the velocity splitting of symmetric matches, but in addition to that the data show a 4$\sigma$ 
excess of pairs relative to random, near zero velocity splitting ($\Delta v < 150$ \kms).  This shows that multiple sightline 
coherence techniques work using line counting when applied to high redshift quasar pairs, and produce 
significant detections of coherence on $\sim$ 2.5 $h^{-1}_{70}$ Mpc scales up to $z \sim 3.8$.

This data set have provided a unique opportunity to test expecations from simulations. The nearest-neighbor
absorber matching, transmission distribution and transverse coherence all indicate agreement between 
simulation expectations and results from data, which has not been reliably tested before at this high 
redshift and wide separation, primarily since quasar pairs of this type are rare. The absorber matching 
experiment shows that the simulations show a weak but detectable coherence signal at low velocity splitting,
with the same 4$\sigma$ strength as the data.  Combining sightline statistics, we compared transmission 
distributions in five discrete bins and see that the data generally agree with the redshift evolution of 
the simulation's transmission distribution shapes, and agree best with the \citet{kirkman05a} treatment 
of mean flux rescaling.  The cross-correlation experiment finds a weak coherence signal from the simulations 
for both mean flux decrements \citep{press93a,kirkman05a}; the data also show a weak signal when rescaled by 
\citet{kirkman05a}, agreeing with the simulation result, but show a strong coherence measurement if rescaled 
by \citet*{press93a}.  These three separate statistics (arbsorber matching, transmission distribution shape 
and cross-correlation amplitude) are mutually consistent and have opened a new regime in redshift and wide 
separation on which to compare the structure of the IGM, and agreement between data and simulations.

\acknowledgements
This work could not have been completed without the help of the excellent day and nighttime staff of the 
Keck Observatory.  We thank Volker Springel and Lars Hernquist for providing us with the G6 simulation 
run, and we thank Jill Bechtold, Jason Prochaska, and Joe Hennawi for helpful advice and conversations 
along the way.  We also thank our referee, John Stocke, for his insightful comments which have helped 
improve the paper.



\end{document}

%% file: tab1.tex
\begin{deluxetable*}{crrrrrrc}
\tabletypesize{\footnotesize}
\tablecaption{Absorption Lines for PC1634+4631A}
\tablewidth{0pt}
\tablehead{
\colhead{Line} &
\colhead{$\lambda_{c}$} &
\colhead{$W_{obs}$} &
\colhead{$\chi_{\nu}^{2}$}  &
\colhead{$S_{fit}$} &
\colhead{$S_{det}$} &
\colhead{$z_{abs}$} &
\colhead{Notes} \\
\colhead{} &
\colhead{(\AA)} &
\colhead{(\AA)} &
\colhead{} &
\colhead{} &
\colhead{} &
\colhead{} &
\colhead{} \\
\colhead{(1)} &
\colhead{(2)} &
\colhead{(3)} &
\colhead{(4)} &
\colhead{(5)} &
\colhead{(6)} &
\colhead{(7)} &
\colhead{(8)} \\
}
\startdata
1 & $4383.28\pm 0.10$ & $ 2.89\pm 0.13$ &  6.07 & 18.40 & 21.42 & 2.6057 & a \\ 
2 & $4390.95\pm 0.21$ & $ 2.12\pm 0.19$ &  1.69 & 14.45 & 11.28 & 2.6120 & a \\ 
3 & $4394.30\pm 0.61$ & $ 0.80\pm 0.19$ &  1.69 &  5.61 &  4.21 & 2.6147 & a \\ 
4 & $4399.17\pm 0.21$ & $ 1.52\pm 0.13$ &  1.69 & 10.94 & 11.50 & 2.6187 & a \\ 
... & & & & & & & \\
240 & $6021.08\pm 0.52$ & $ 0.31\pm 0.07$ &  2.73 &  4.45 &  4.40 & \nodata & c \\ 
\enddata
\label{tab1}
\tablecomments{
For each absorption feature listed in col. (1), col. (2) gives the central 
wavelength in \angns, and col. (3) gives the equivalent width in \angns.
The reduced $\chi^{2}$ is listed in col. (4), and if this value is identical
to that of an adjacent line, it indicates that the lines were fitted 
simultaneously. Col. (5) gives the significance of the line defined as $W_{obs}/\sigma_{fit}$, 
where $\sigma_{fit}$ is the error in the equivalent width, and col. (6) gives 
the significance of the line defined as $W_{obs}/\sigma_{det}$, where 
$\sigma_{det}$ is the detection limit of the data at $\lambda_{c}$ (see text for further 
details). Cols. (7) and (8) contain the redshift of the absorber associated with
its identification.  
The Notes labels are as follows: (a) lines included in the \lya sample, (b) 
lines associated with the opaque system at $z = 3.777$ that are also labeled by their 
associated identification, (c) lines redward of \lya emission which are clearly 
excluded from the \lya sample. [The complete version of this table may be found in the
electronic edition of the journal]
}
\end{deluxetable*}

%% file: tab2.tex
\begin{deluxetable*}{crrrrrrc}
\tabletypesize{\footnotesize}
\tablecaption{Absorption Lines for PC1634+4631B}
\tablewidth{0pt}
\tablehead{
\colhead{Line} &
\colhead{$\lambda_{c}$} &
\colhead{$W_{obs}$} &
\colhead{$\chi_{\nu}^{2}$}  &
\colhead{$S_{fit}$} &
\colhead{$S_{det}$} &
\colhead{$z_{abs}$} &
\colhead{Notes}\\
\colhead{} &
\colhead{(\AA)} &
\colhead{(\AA)} &
\colhead{} &
\colhead{} &
\colhead{} &
\colhead{} &
\colhead{} \\
\colhead{(1)} &
\colhead{(2)} &
\colhead{(3)} &
\colhead{(4)} &
\colhead{(5)} &
\colhead{(6)} &
\colhead{(7)} &
\colhead{(8)} \\
}
\startdata
1 & $4428.96\pm 0.38$ & $ 2.24\pm 0.39$ &  0.38 &  5.45 &  5.77 & 2.6432 & a \\ 
2 & $4438.54\pm 0.35$ & $ 2.23\pm 0.33$ &  0.30 &  6.66 &  6.74 & 2.6511 & a \\ 
3 & $4443.19\pm 0.30$ & $ 2.99\pm 0.34$ &  0.30 &  9.17 &  8.88 & 2.6549 & a \\ 
4 & $4447.21\pm 0.29$ & $ 2.80\pm 0.33$ &  0.30 &  8.80 &  8.49 & 2.6582 & a \\ 
... & & & & & & & \\
234 & $6001.46\pm 0.31$ & $ 1.25\pm 0.20$ &  0.36 & 13.30 &  6.27 & \nodata & e \\ 
\enddata
\label{tab2}
\tablecomments{
Same as Table~1, but for quasar PC1643+4631B.
}
\end{deluxetable*}

%% file: tab3.tex
\begin{deluxetable*}{crrrrrrr}
\tabletypesize{\footnotesize}
\tablecaption{Symmetrically Matched \lya Lines}
\tablewidth{0pt}
\tablehead{
\colhead{Match Number} &
\colhead{$\lambda_{A}$} &
\colhead{$\lambda_{B}$} &
\colhead{$|\Delta\lambda|$} &
\colhead{$|\Delta v|$} &
\colhead{$W_{0,A}$} &
\colhead{$W_{0,B}$} &
\colhead{$|\Delta W_{0}|$} \\
\colhead{} &
\colhead{(\AA)} &
\colhead{(\AA)} &
\colhead{(\AA)} &
\colhead{(\kms)} &
\colhead{(\AA)} &
\colhead{(\AA)} &
\colhead{(\AA)} \\
\colhead{(1)} &
\colhead{(2)} &
\colhead{(3)} &
\colhead{(4)} &
\colhead{(5)} &
\colhead{(6)} &
\colhead{(7)} &
\colhead{(8)} \\
}
\startdata

      1  &    4428.00  &    4428.95   &    0.95   &   65.0   &    0.08   &    0.61   &    0.52  \\
      2  &    4439.29  &    4438.54   &    0.75   &   50.6   &    0.67   &    0.60   &    0.06  \\
      3  &    4445.02  &    4443.18   &    1.83   &  123.5   &    0.93   &    0.81   &    0.11  \\
      4  &    4449.20  &    4447.20   &    2.00   &  134.8   &    0.38   &    0.76   &    0.37  \\
... & & & & & & & \\
    152  &    5781.83  &    5781.31   &    0.52   &   27.4   &    0.25   &    0.37   &    0.11  \\
\enddata
\label{tab3}
\tablecomments{
  Each symmetrically matched pair is labeled 1-152 in col. (1), with col. (2) and (3) giving the wavelengths
  of the matched absorbers in spectrum A and B respectively.  Column (4) gives the wavelength separation of the
  pair, and col. (5) gives the velocity difference in \kms.  Columns (6) and (7) give the rest equivalent widths of
  the lines (for A and B respectively) and col. (8) gives the difference in rest equivalent widths.
  [The complete version of this table may be found in the electronic edition of the Journal]
}
\end{deluxetable*}